\documentclass[article,10pt]{IEEEtran}

\usepackage[latin1]{inputenc}
\usepackage{xcolor}
\usepackage{cite}
\usepackage{amssymb}
\usepackage{amsthm,amsmath,array} 
\usepackage{bigints}              
\usepackage{rotating}
\usepackage{bm}
\usepackage{dsfont}
\usepackage{acronym}
\usepackage{enumitem}
\usepackage{cleveref}
\usepackage{tabularx}
\usepackage{mathrsfs}
\usepackage[caption=false,font=footnotesize]{subfig}
\crefname{equation}{}{} 
\creflabelformat{equation}{(#2#1#3)}
\usepackage{footmisc}

\newcommand{\ist}{\hspace*{.3mm}}
\newcommand{\rmv}{\hspace*{-.3mm}}
\newcommand{\iist}{\hspace*{1mm}}
\newcommand{\rrmv}{\hspace*{-1mm}}
\newcommand{\nn}{\nonumber}

\usepackage{bigstrut}
\usepackage{tabularx}
\usepackage[caption=false,font=footnotesize]{subfig}
\usepackage{pgfplots}
\usepackage{tikzscale}
\newcommand{\exportFigures}{true} 

\usepackage[ruled,vlined,linesnumbered]{algorithm2e}
\SetAlFnt{\small}
\SetAlCapFnt{\normalsize}
\SetAlCapNameFnt{\normalsize}
\usepackage{algorithmic}
\algsetup{linenosize=\tiny}

\SetKwRepeat{Do}{do}{while}
\makeatletter
\newcommand{\RemoveAlgoNumber}{\renewcommand{\fnum@algocf}{\AlCapSty{\AlCapFnt\algorithmcfname}}}
\newcommand{\RevertAlgoNumber}{\algocf@resetfnum}
\makeatother

\newcommand{\tikzfolder}{./compiledPlots/}

\input{definitions}

\acrodef{mimo}[MIMO]{multiple-input---multiple-output}
\acrodef{simo}[SIMO]{single-input---multiple-output}
\acrodef{sc}[SC]{specular component}
\acrodef{dmc}[DMC]{dense multipath component}
\acrodef{dc}[DC]{dense component}
\acrodef{aoa}[angle]{angle-of-arrival}
\acrodef{los}[LOS]{line-of-sight}
\acrodef{uwb}[UWB]{ultra-wide-band}
\acrodef{tx}[Tx]{transmitter}
\acrodef{rx}[Rx]{receiver}
\acrodef{mmse}[MMSE]{minimum mean-square error}
\acrodef{rmse}[RMSE]{root mean-square error}
\acrodef{ospa}[OSPA]{optimal subpattern assignment}
\acrodef{mospa}[MOSPA]{mean optimal subpattern assignment}
\acrodef{pdf}[PDF]{probability density function}
\acrodef{sbl}[SBL]{sparse Bayesian learning}
\acrodef{ml}[ML]{maximum-likelihood}
\acrodef{mle}[MLE]{maximum-likelihood estimation}
\acrodef{wgn}[WGN]{white Gaussian noise}
\acrodef{awgn}[AWGN]{additive \ac{wgn}}
\acrodef{snr}[SNR]{signal-to-noise-ratio}
\acrodef{map}[MAP]{maximum-a-posteriori}
\acrodef{ssr}[SSR]{sparse signal reconstruction}
\acrodef{dps}[DPS]{delay power spectrum}
\acrodef{aps}[APS]{angle power spectrum}
\acrodef{daps}[DAPS]{delay-angle power spectrum}
\acrodef{crb}[CRLB]{Cram\'er-Rao lower bound}
\acrodef{isac}[ISAC]{integrated sensing and communications}
\acrodef{slam}[SLAM]{simultaneous localization and mapping}
\acrodef{vna}[VNA]{vector network analyzer}
\acrodef{em}[EM]{{E}xpectation-maximization}
\acrodef{mdl}[MDL]{minimum description length}
\acrodef{bic}[BIC]{Bayesian information criterion}
\acrodef{aic}[AIC]{Akaike information criterion}
\acrodef{atm}[ATM]{atomic noise minimization}
\acrodef{lse}[LSE]{line spectral estimation}
\acrodef{rrl}[RRL]{Rayleigh resolution limit}

\IEEEoverridecommandlockouts
\begin{document}
	\allowdisplaybreaks
	\frenchspacing
   	\title{\huge Super-Resolution Estimation of UWB Channels including the Diffuse Component --- An SBL-Inspired Approach}
	\author{\normalsize Stefan Grebien, 
		Erik Leitinger~\IEEEmembership{\normalsize Member,~IEEE}, 
		Klaus Witrisal~\IEEEmembership{\normalsize Member,~IEEE},
		and Bernard H. Fleury~\IEEEmembership{\normalsize Senior Member,~IEEE}\\[1.5mm]
		\vspace*{-10mm}
		\thanks{S.\ Grebien, E.\ Leitinger, and K.\ Witrisal are with the Laboratory of Signal Processing and Speech Communication, Graz University of Technology, Graz, Austria, and Christian Doppler Laboratory for Location-aware Electronic Systems (e-mail: \{stefan.grebien, erik.leitinger, witrisal\}@tugraz.at). 
		B.\ Fleury is with the Institute of Telecommunications, Vienna University of Technology, Vienna, Austria, (bernard.fleury@tuwien.ac.at). S.\ Grebien and E.\ Leitinger have equally contributed as first authors.
		}\vspace{0mm}}
	
	\maketitle
	\renewcommand{\baselinestretch}{0.95}\small\normalsize
	\begin{abstract}
		In this paper, we present an iterative algorithm that detects and estimates the specular components and estimates the diffuse component of \ac{simo} \ac{uwb} multipath channels. Specifically, the algorithm super-resolves the specular components in the delay\---angle-of-arrival domain and estimates the parameters of a parametric model of the delay-angle power spectrum characterizing the diffuse component. Channel noise is also estimated. In essence, the algorithm solves the problem of estimating spectral lines (the specular components) in colored noise (generated by the diffuse component and channel noise). Its design is inspired by the \ac{sbl} framework. As a result the iteration process contains a threshold condition that determines whether a candidate specular component shall be retained or pruned. By relying to results from extreme-value analysis the threshold of this condition is suitably adapted to ensure a prescribed probability of detecting spurious specular components. Studies using synthetic and real channel measurement data demonstrate the virtues of the algorithm: it is able to still detect and accurately estimate specular components, even when their separation in delay and angle is down to half the \ac{rrl} of the equipment; it is robust in the sense that it tends to return no more specular components than the actual ones. Finally, the algorithm is shown to outperform a state-of-the-art super-resolution channel estimator.

	\end{abstract}                      
	\renewcommand{\baselinestretch}{0.978}\small\normalsize
	\vspace*{-4mm}
	\section{Introduction}\label{sec:Introduction}
	Future wireless communication technologies will support a variety of services with high quality requirements, addressing performance metrics such as reliability, ultra-low latency, high data rates, and resource-efficient use of the infrastructure \cite{ConMorLiuBarMazLinWin:CM2021,Wymetal:EuCNC2022}. Holistic approaches that combine different functionalities have proven to offer promising solutions to meet these requirements. Illustrative examples are \ac{isac} and radio-based \ac{slam}  \cite{KwoConParWin:JSTSP2021,BjoeSanWymHoyMar:DSP2019,Wymetal:EuCNC2022}. These examples emphasize the reliance of these technologies on extended, accurate channel state information. High-performance feasible parametric multi-antenna channel estimators can provide this information.

\vspace*{-2mm}
\subsection{State of the Art}

Parametric channel models typically represent multipath propagation as a linear superposition of weighted Dirac delta distributions \-- or spectral lines \-- with distinct supports in the underlying dispersion domain (delay, angle of arrival, angle of departure, Doppler frequency, and combinations thereof). Each component in the superposition is meant to represent a \acf{sc}. Note that in this paper we shall use the terms \ac{sc} and spectral line indiscriminately. The finite aperture of the measurement equipment imposes some limitation on the ability to resolve \acp{sc} closely spaced in the dispersion domain.

If the number of spectral lines is known, (constrained and unconstrained) \acf{ml} methods, see e.g. \cite{Ottersten1993} or subspace-based methods \cite{Schmidt1986,Roy1989} are standard super-resolution\footnote{Super-resolution is the ability of an algorithm to resolve spectral lines even if the separation of their support in the dispersion domain is below the intrinsic resolution of the measurement equipment.} tools to estimate their parameters. \acl{em} and related algorithms \cite{Fleury1999,RichterPhD2005}
have proven viable approximations of the computationally prohibitive direct implementation of the constrained \ac{ml} method. These estimators have in common that they do not incorporate the estimation of the number of spectral lines into the estimation problem. Schemes that perform jointly detection of the spectral lines and estimation of their parameters have been designed within a Bayesian framework \cite{Dublanchet1997, AndrieuTSP1999}. Traditional methods combining detection and estimation select among multiple candidate models, each corresponding to a specific hypothesis on the number of spectral lines, the one that optimizes a so-called information criterion, such as the Akaike or Bayesian information criterion, and the \acl{mdl}, see \cite{StoicaTSP2004} and references therein. Yet, the information-based approach suffers from two shortcomings: (a) it is computationally intensive as the adopted information criterion needs to be computed first for each model candidate before a decision can be made; (b) the number of spectral lines of the selected model tends to be positively biased in non-asymptotic regimes of the \ac{snr} and the number of observed samples \cite{DjuricTSP1996}. Hence, inference schemes designed with this approach are prone to return spurious spectral lines that have no real counterpart. Alternative penalty terms have been proposed that prevent \cite{NadlerTSP2011} or control \cite{MarianiTSP2015} this bias.

Model-order selection is inherently realized in \ac{ssr}, see  \cite{TroppProc2010} and references therein. 
\ac{ssr} aims at recovering a sparse weight vector in an underdetermined linear model with a known and fixed dictionary matrix.  To that end it computes an estimate of the weights as the solution to a regularized optimization problem in which the regularization term is selected to promote sparse solutions. A popular instance of \ac{ssr} is basis pursuit denoising \cite{ChenSIAM2001}, also called LASSO (least absolute shrinkage and selection operator) \cite{TibshiraniJRSS1996}, that uses an $\ell_1$-norm regularization. \ac{ssr} can be formulated within the Bayesian framework as \ac{map} estimation while imposing a sparsity promoting prior on the weight vector. Typically this prior is endowed with a hierarchical structure involving a hyperparameter for each weight. Several hierarchical models have been considered so far: gamma-Gaussian\footnote{This Bayesian formulation with this choice of hierarchical model is also referred to as sparse Bayesian learning or relevance vector machine to stress its link with automatic relevance determination \cite{MacKay1996} that uses a similar hierarchical model.} \cite{TippingJMLR2001, WipfTSP2004}, Bernoulli-Gaussian \cite{Champagnat1996, Soussen2011},  
and generalized-gamma\---power-exponential \cite{GiriTSP2016}. This Bayesian formulation has proven to be a particularly flexible and effective tool for \ac{ssr}. 
Since direct implementation of the estimators is typically computationally prohibitive, one has to resort to iterative schemes, often designed using variational inference methods \cite{Bishop2000UAI,TzikasRTSP2008}.

\ac{ssr} can be straightforwardly applied in the context of line spectral estimation by discretizing (gridding) the dispersion domain, see e.g. \cite{Stoica2011,Tan2011,PoteTSP2023}. The benefit of doing so is that the complex optimization problem needed to estimate the supports of the spectral lines is replaced by a linear programming procedure that returns a sparse estimate of the weight vector. The shortcoming is that gridding of the dispersion domain induces spectral leakage due to the resulting model mismatch. This effect can be mitigated by selecting a denser grid, yet at the cost of increasing the coherence of the dictionary matrix, which impairs the sparse reconstruction capability and increases the computational complexity. Variants of gridding methods that employ some interpolation method \cite{Ekanadham2011,Yang2013,Duarte2013,Hu2013,Fyhn2015} or apply a grid refinement technique \cite{MalioutovTSP2005, PoteTSP2023} have been proposed to circumvent the leakage effect. 

\Ac{atm} provides an elegant natural means to operate with a continuous, i.e. infinite, dictionary in \ac{ssr} and thereby to relax the need for discretizing the dispersion domain 
\cite{BhaskarTSP2013,YanXie:TSP2015,ChiTSP2015}. However, some specificities of our underlying model --- namely a two-dimensional dispersion domain and unknown colored noise --- prevent a direct application of the method, see Subsection~\ref{s:Discussion} and a related discussion in \cite{ChiTSP2015}. Moreover, numerical evidence shows that \ac{atm} requires the supports of spectral lines to be sufficiently separated in the dispersion domain in order to be able to recover them \cite{YanXie:TSP2015}. In \cite{Yang2016} an alternative is proposed that circumvents this shortcoming.

In theory, gridding-based line spectral estimation methods can be straightforwardly extended to account for continuous dispersion parameters by relaxing the discretization constraint and instead including the estimation of the support of the spectral lines in the inference process. Clearly, this approach is an instance of \ac{ssr} with learning the continuous (vector-valued) parameter of a parameterized dictionary matrix. It has been extensively pursued in connection with the Bayesian formulation of \ac{ssr} \cite{ShutinTSP2011,ShutinCSTA2013,Hansen2013,HansenSAM2014, ShutinArxiv2015, HansenTSP2018, BadiuTSP2017}. These algorithms differ in their specific design criteria, such as (i) the chosen sparsity-inducing hierarchical prior model, e.g. gamma-Gaussian \cite{ShutinTSP2011,ShutinCSTA2013,Hansen2013,HansenSAM2014,ShutinArxiv2015}, Bernoulli-Gaussian \cite{BadiuTSP2017, HansenTSP2018}, (ii) the assumed absence \cite{ShutinTSP2011, BadiuTSP2017} or presence \cite{ShutinCSTA2013,ShutinArxiv2015} of correlation among the weights of the spectral lines, and (iii) whether point estimates \cite{ShutinTSP2011,Hansen2013,HansenSAM2014, ShutinArxiv2015,HansenTSP2018} or posterior \acp{pdf} of the dispersion parameters of the \acp{sc} are inferred \cite{BadiuTSP2017}. Experimental evidence shows that the algorithms computing point estimates of the supports of spectral lines show a positive bias in the number of detected spectral lines, i.e. are prone to detect spurious spectral lines. Including inference of the posterior \ac{pdf} of the supports allows for mitigating this bias, yet at the cost of an increased computational complexity \cite{BadiuTSP2017}. 
We remark that the previously mentioned (iterative) \ac{ssr} methods that apply grid refinement techniques \cite{MalioutovTSP2005, PoteTSP2023} can be viewed as particular instances of \ac{ssr} methods with continuous-parameter learning, which adapt their inherent restricted range of the dictionary parameter during the iterations.\footnote{For instance, the ``gridless'' SBL-based method presented in \cite[Sec.~IV]{PoteTSP2023} is similar to the methods proposed in \cite{Hansen2013,HansenSAM2014}.}

The above \ac{ssr} methods with continuous-parameter dictionary learning include an inherent pruning procedure that determines which ones among the columns of the dictionary matrix are inferred as relevant and switch the others off, see e.g. \cite{Tipping2003,ShutinTSP2012}. 
It is shown in \cite{ShutinArxiv2015} that the number of detected spurious spectral lines 
can be significantly reduced by suitably adapting the threshold of the pruning stage. 
The analysis provided there relies on some heuristic, yet realistic, assumptions that allow for approximating the probability of detecting a spurious line with the probability that the maximum of a continuous $\chi^2$ random field exceeds the selected threshold \cite{WorsleyAIAP1994,Adler2015ApplicationsRandomFieldsGeometry}. 
The analysis shows that a prescribed probability of detecting spurious lines can be guaranteed, provided the threshold increases as $C+\log{n}+\frac{1}{2}\log{\log{n}}$ where $n$ is the number of observation samples and $C$ is a constant that depends on that probability \cite{NadlerTSP2011,ShutinArxiv2015}. 
Numerical analyses have shown that using this adapted threshold leads to almost vanishing bias in the number of detected SCs in medium and high \ac{snr} regimes with a tendency to underestimate said number in the low \ac{snr} regime, see also Section~\ref{sec:results}.

In recent years, an extension of the channel model has been considered, that includes a \ac{dc} \cite{RichterPhD2005}. The \ac{dc} incorporates diffuse components as well as \acp{sc} that cannot be resolved with the finite aperture of the measurement equipment. Including the estimation of the \ac{dc} can improve the accuracy of the estimation of the parameters of resolved \acp{sc} \cite{RichterPhD2005}.
\vspace*{-2mm}

\subsection{Contributions of the Paper}

We propose an iterative algorithm that performs combined detection and estimation of \acp{sc} and estimation of the \ac{dc} plus \ac{awgn} in \acf{simo} \acf{uwb} multipath channels.\footnote{The extension of the algorithm to a \ac{mimo} system is straightforward \cite{LeitingerAsilomar2020}.} 
%
%
The algorithm resolves the \acp{sc} in the delay\---\ac{aoa} domain. 
The contributions of this paper are as follows:
\begin{itemize}
  \item  We model the impact of the \ac{dc} and AWGN as a colored noise, so that the problem becomes that of line spectral estimation \cite{stoica2005spectral} in such noise when the relative delays that the (\ac{uwb}) complex envelope of the sounding wave exhibits when it is sensed by the elements of the antenna array cannot be neglected.
  \item 
The design of the algorithm is inspired by the SBL approach \cite{TippingJMLR2001}. The probabilistic model is extended by assuming that the weights of the spectral lines are independent circularly-symmetric complex Gaussian random variables with unknown variances. In a first stage \ac{ml} estimation of the variances and all other parameters but the weights is performed after integrating out said weights. These estimates are then used to compute a tractable (Gaussian) approximation of the weights' posterior \ac{pdf}. The algorithm computes these two stage, the former one in an iterative fashion.
 \item We suitably modify the threshold inherent to the above \ac{ml} estimation stage 
 to meet a prescribed probability of detecting spurious lines. To do so we apply results from extreme value analysis \cite{WorsleyAIAP1994, Adler2015ApplicationsRandomFieldsGeometry}.
  \item Using synthetically generated observation data we study in-depth the behavior of the proposed algorithm and especially how the adapted threshold affects its performance.
  \item We compare the performance of the algorithm with that of a state-of-the-art combined detection and estimation scheme that relies on the information criterion derived in  \cite{NadlerTSP2011}.
  \item We apply the algorithm to \ac{uwb} measurement data collected in an indoor environment. A simple ray-tracing tool is used to identify plausible propagation paths that can be associated to the \acp{sc} detected by the algorithm.
\end{itemize}

The remainder of the paper is organized as follows: In Section~\ref{sec:signalmodel} we present the generative signal model for the considered \ac{simo} measurement set-up. 
Section~\ref{sec:problemformulation} describes the probabilistic signal model for inference.  
We derive the proposed algorithm in Section~\ref{sec:SBVI_algorithm}. 
Section~\ref{sec:analyPrun} addresses the analytical correspondence between the probability of detecting spurious spectral lines and the threshold of the \ac{ml} estimation stage. 
Section~\ref{sec:results} reports results from numerical and experimental studies. 
Concluding remarks are provided in Section~\ref{sec:concl}.

	\section{Signal Model}\label{sec:signalmodel}
	\subsection{Continous-Time Signal Model}\label{sec:rsigmod}

\begin{figure}[t!]
\vspace*{-5mm}
\centering
\scalebox{0.8}{
\includegraphics{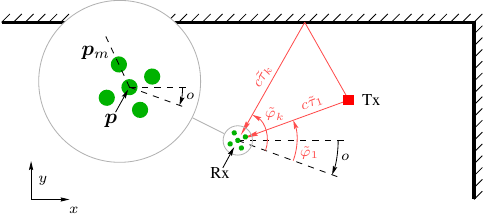}}
\vspace*{-3mm}
\caption{Layout of the array with its center of gravity $\V{p}$, the $m$th element position $\V{p}_m$ and reference orientation $o$. The $1$th  \ac{sc} originates from propagation along the direct path from the \ac{tx} to the \ac{rx} with \ac{aoa} $\tilde{\varphi}_1$ and path distance $c\tilde{\tau}_1$. The $k$th \ac{sc} is incident with angle $\tilde{\varphi}_k$ and path distance $c\tilde{\tau}_k $.}
\vspace*{-3mm}
\label{fig:geometry}
\end{figure}

The experimental measurement setup consists of an \ac{uwb} transceiver operating in an indoor environment. The \ac{tx} is equipped with a single antenna, while an antenna array with colocated elements is emulated at the \ac{rx} using a single antenna mounted on a positioning table. For the sake of simplicity we assume horizontal-only propagation.\footnote{An extension to three dimensional propagation scenarios including polarization is straightforward, but more involved.} The array at the \ac{rx} has $M$ elements located at $\V{p}_m \in \mathbb{R}^2$, $m \in \{1,\dots,M\} \triangleq \mathcal{M}$, see Fig.~\ref{fig:geometry}. Its center of gravity is $\V{p}=M^{-1}\sum_{m=1}^M\V{p}_m$ and its orientation determined by the angle $o$ as depicted in the figure. 

Signals are represented by means of their complex envelope with respect to a center frequency $f_\mathrm{c}$. Under the plane-wave assumption, 
the signal at the output of the $m$th antenna element reads 
\vspace*{-1mm}
\begin{align}\label{eq:tc_signal_model}
  r_m(t)\,\, &= 
\int\rmv\rmv\rmv\rmv\rmv\int  s \big(t;\tau,\varphi,\V{p}_m\big) h(\tau,\varphi) \textrm{d}\tau \textrm{d} \varphi + w_m(t)\ist .\\[-6mm]\nn
\end{align} 
In this expression
\vspace*{-1mm}
\begin{align}\label{eq:s_signal_time}
s\big(t;\tau,\varphi,\V{p}_m\big) = \text{e}^{j2\pi f_c g(\varphi,\V{p}_m)} \underline{s}(t-(\tau - g(\varphi,\V{p}_m)))\ist
 \\[-6mm]\nn
 \end{align}
where $\underline{s}(t)$ is the transmitted signal with bandwidth $B$ and $g\big(\varphi, \V{p}_m\big) \rmv\rmv=\rmv\rmv [\cos(\varphi)\iist\sin(\varphi)](\V{p}_m-\V{p})/{c}$ with $c$ denoting the speed of light, 
expresses for a plane wave incident with \ac{aoa} $\varphi\in[-\pi,+\pi)$ the wave's excess (propagation) delay at $\V{p}_m$ relative to the reference point $\V{p}$.
%
The function $h(\tau,\varphi)\in\mathbb{C}$ defined on $\in\mathbb{R}\times[-\pi,+\pi)$
characterizes the spread in (relative) delay $\tau$ and \ac{aoa} $\varphi$ 
of the signal sensed at $\V{p}$. Finally, $w_m(t)$, $m\rmv\rmv\in\rmv\rmv\mathcal{M}$ are independent \ac{wgn} with double-sided power spectral density $N_0/2$.

We see from \eqref{eq:tc_signal_model} that sufficient conditions for this identity to be accurate are that (a) the plane wave assumption holds over the \ac{rx} array aperture (determined by $\{\V{p}_1,\ldots, \V{p}_m\}$)\footnote{Strictly speaking the aperture of the virtual array that we emulated in this study also incorporates the radiation pattern of the used antenna.}, i.e. the array is located far away enough from the \ac{tx} and the objects in the environment that notably contribute to multipath propagation, such as walls, boards, etc., and (b) that the spread function $h(\tau,\varphi)$ stays constant over the bandwidth (frequency aperture) of the sounding signal. The latter assumption implies that the electromagnetic properties of said objects, like reflection and transmission coefficients,  are nearly constant over the sounding bandwidth.

In this study we assume that the delay-\ac{aoa} spread function $h(\tau,\varphi)$ is the sum of the superposition of a finite number, say $K$, of spectral lines representing \acp{sc} and a (spread) function $\nu(\tau,\varphi)$ describing the \ac{dc}, i.e.
\vspace*{-1mm}
\begin{align}\label{eq:channel}
  h(\tau,\varphi) &= \sum_{k \in \mathcal{K}} \tilde{\alpha}_{k} \delta(\tau-\tilde{\tau}_{k})\delta(\varphi - \tilde{\varphi}_k) + \nu(\tau,\varphi)\\[-7mm]\nn
\end{align}
where $\delta(\cdot)$ denotes the Dirac delta distribution. The $k$th \ac{sc}, $k \in \mathcal{K}\triangleq\{1,\dots,K\}$ is characterized by its complex amplitude $\tilde{\alpha}_k\in \mathbb{C}$, its (relative) delay $\tilde{\tau}_k\in \mathbb{R} $ and \ac{aoa} (of arrival) $\tilde{\varphi}_k\in[-\pi,+\pi)$. 
We model $\nu(\tau,\varphi)$  as a complex circular symmetric (i.e. zero-mean) Gaussian random process \cite{RichterPhD2005,KaredalTWC2007}. 
Furthermore, we assume uncorrelated scattering, i.e.,
$\mathbb{E}[\nu(\tau',\varphi')\nu^{*}(\tau,\varphi)] = P(\tau,\varphi)\delta(\tau'-\tau)\delta(\varphi' - \varphi)$ \cite{FleuryTIT2000}, where $\mathbb{E}[\cdot]$ denotes expectation and $P(\tau,\varphi)$ is the \ac{daps} of the \ac{dc} \cite{FleuryTIT2000}. 
We make the following additional hypotheses: (a) The 
spread function $h(\tau,\varphi)$ has bounded support, i.e., without loss of generality, $h(\tau,\varphi)=0$ whenever $[\tau\iist\iist\varphi] \notin[0,T)\times[-\pi,\pi)=\Psi$ with $T>0$; (b) the equipment is designed in such a way to ensure an aliasing-free estimation of $h(\tau,\varphi)$ over $\Psi$. Condition (a) implies that $P(\tau,\varphi;\V{\vartheta})=0$ whenever $[\tau\iist\iist\varphi]\notin\Psi$. It also imposes that the dispersion vector $[\tilde{\tau}_k\iist\iist\tilde{\varphi}_k]$ of any $k$th \ac{sc}, $k\in\mathcal{K}$ belongs to the dispersion domain $\Psi$.

Inserting the decomposition \eqref{eq:channel}  in \eqref{eq:tc_signal_model} yields
\vspace*{-1mm}
\begin{align}\label{eq:rx_signal}
	r_m(t)\,\, &= \sum_{k \in \mathcal{K}}\tilde{\alpha}_{k} s\big(t; \tilde{\tau}_k, \tilde{\varphi}_k,\V{p}_m\big)  \nn\\
	&\hspace*{4mm}+ \int\rmv\rmv\rmv\rmv\rmv\int  s \big(t;\tau,\varphi,\V{p}_m\big) \nu(\tau,\varphi) \textrm{d}\tau \textrm{d} \varphi + w_m(t)\ist .\\[-6mm]\nn
\end{align}
The rationale behind the selection of model \eqref{eq:channel} is as follows. The \acp{sc} originate from electromagnetic interactions with objects in the environment that are essentially non-dispersive, such as \ac{los} propagation, specular reflection and transmission, and can be resolved with the used aperture. The \ac{dc} incorporates the contributions from all other interactions, e.g. diffuse scattering and diffraction. It also includes components from specular interactions that cannot be resolved with the used aperture.

\subsection{Discrete-Frequency Signal Model}\label{sec:gensigmod}
The signals $r_m(t)$, $m\in\mathcal{M}$ are Nyquist filtered, Fourier transformed, and then synchronously and uniformly sampled with frequency spacing $\Delta$ over the bandwidth $B$ to collect for each branch $m$ $N=B/\Delta$ samples that are arranged in a $N$-dim. vector $\V{y}_m$. These $M$ vectors are then stacked to form the $NM$-vector $\V{y}\rmv=\rmv [\V{y}_1^{\ist\text{T}}\iist\cdots\iist \V{y}_M^{\ist\text{T}}]\trans$, which is expressed as
\vspace*{-1mm}
\begin{align}\label{eq:stack_recvgen}
 \V{y} = \V{S}(\tilde{\V{\psi}})\tilde{\V{\alpha}} + \V{n} \quad \in \mathbb{C}^{NM \times 1}\ist \\[-6mm]\nn
\end{align}
with $\tilde{\V{\alpha}} = [\tilde{\alpha}_1 \iist\cdots\iist \tilde{\alpha}_{K}]\trans \in \mathbb{C}^{K\times 1}$, 
$\tilde{\V{\psi}} = [\tilde{\V{\psi}}_1\iist\cdots\iist \tilde{\V{\psi}}_K]\in\Psi^{K}$, 
and $\V{S}(\tilde{\V{\psi}})\rmv\rmv=\rmv\rmv[\V{s}(\tilde{\V{\psi}}_1) \ist\cdots\ist \V{s}(\tilde{\V{\psi}}_{K})] \rmv\in\rmv \mathbb{C}^{NM\times K}$ with columns given by
\vspace*{-1mm}
\begin{align}\label{eq:vs}
\V{s}(\tilde{\V{\psi}}_k) &= [\V{s}_{1}(\tilde{\V{\psi}}_k)\trans\iist \cdots \iist \V{s}_M(\tilde{\V{\psi}}_k)\trans]\trans \in \mathbb{C}^{NM\times1}\ist,\iist k\in\mathcal{K}\ist.\\[-6mm]\nn
\end{align}
With $S(f;\tau, \varphi, \V{p}_m)$ denoting the Fourier transform of $s(t;\tau, \varphi, \V{p}_m)$, i.e., 
\vspace*{-1mm}
\begin{align}\label{eq:s_signal}
	S\big(f;\tau,\varphi,\V{p}_m\big) = \text{e}^{j2\pi f_c g(\varphi,\V{p}_m)} \underline{S}(f) \text{e}^{-j2\pi f(\tau - g(\varphi,\V{p}_m))}\ist
	\\[-6mm]\nn
\end{align} 
where $\underline{S}(f)$ is the Fourier spectrum of $\underline{s}(t)$, the $m$th entry 
in the vector in \eqref{eq:vs} reads 
\vspace*{-1mm}
\begin{align}\label{eq:sm}
\V{s}_m(\tilde{\V{\psi}}_k) &\triangleq\big  [S\big(n\Delta;\tilde{\tau}_k, \tilde{\varphi}_k,\V{p}_m\big): n \nn\\
&\hspace*{3mm}=-(N-1)/2,\iist\ldots\iist,(N-1)/2\big]\trans\in \mathbb{C}^{N\times 1}\ist,
\\[-6mm]\nn
\end{align}
i.e., it contains the Fourier-transformed samples collected at the $m$th antenna element.
The $NM$-vector $\V{n} = \V{v} + \V{w}$ 
in \eqref{eq:stack_recvgen} aggregates the vectors $\V{v}$ and $\V{w}$  that collect
the samples (arranged in the right order) corresponding to, respectively, the integral term and the noise term in \eqref{eq:rx_signal} 
when $m$ ranges in $\mathcal{M}$.
From the assumptions on the \ac{dc}, $\V{v}$ is a complex circular symmetric Gaussian random vector with zero mean and $MN\times MN$ covariance matrix $\V{Q}_{\V{v}}=[ [\V{Q}_v]_{m,m'}, \iist m,m' \in \mathcal{M} ]$ with submatrices
\vspace*{-1mm}
\begin{align}\label{eq:C_DMC_full}
[\V{Q}_{\V{v}}]_{m,m'} =\int\rmv\rmv\rmv\rmv\rmv\int \rmv\rmv P(\tau,\varphi) \V{s}_m(\tau,\varphi) \V{s}_{m'}(\tau,\varphi)\herm \mathrm{d}\tau \mathrm{d}\varphi\\[-6mm]\nn
\end{align}
with $(m,m')\in\mathcal{M}^2$. From the assumptions on the noise measurement process $\V{w}$ is a  complex circular symmetric Gaussian random vector with covariance matrix
$\V{Q}_{\V{w}}=\sigma^2 \V{I}_{NM}$ where $\sigma^2=N_\mathrm{0}/T_\mathrm{s}$
and $\V{I}_{(\cdot)}$ is the identity matrix of dimensions specified by the number given in the subscript. 
We assume that $\V{v}$ and  $\V{w}$ are uncorrelated. As a result $\V{n}$ is a circularly symmetric Gaussian random vector with covariance matrix
\vspace*{-1mm}
\begin{align}
\V{Q} = \V{Q}_{\V{v}}+\sigma^2 \V{I}_{NM}\ist.
\label{eq:C-EN}\\[-7mm]\nn
\end{align}

\subsection{Selected Model for the \ac{dc}}\label{ssec:SelectedModel}

We impose some structure on the covariance matrix 
$\V{Q}_{\V{v}}$ in \eqref{eq:C-EN} via some assumptions on the behaviour of the \ac{dc} and simplifying approximations in the derivations of the submatrices in \eqref{eq:C_DMC_full}. This structure will ensure the feasibility of the estimation algorithm.

\textit{a)}\ist\ist\ist\ist The \ac{daps} factorizes as $P(\tau,\varphi) = P\, p(\tau)p(\varphi)$. Here, $P=\int\!\!\int P(\tau,\varphi)\mathrm{d}\tau\mathrm{d}\varphi$ is the power of the \ac{dc}, and $p(\tau)$ and $p(\varphi)$ are, respectively, the normalized \ac{dps} and the normalized \ac{aps}  \cite{AdlerTaylor2007RandomFieldsGeometry}. 

\textit{b)}\ist\ist\ist\ist In the computation of \eqref{eq:C_DMC_full} we discard the second occurrence of the term $g\big(\varphi,\V{p}_m\big)$ in \eqref{eq:s_signal}, i.e., $S(f; \tau, \varphi,\V{p}^{(m)}) = \text{e}^{j2\pi f_c g(\varphi,\V{p}_m)} \underline{S}(f) \text{e}^{-j2\pi f\tau}$. This step amounts to adopting a narrowband representation that neglects the relative delays across array elements of the modulating signals of incident waves \cite{WeichselbergerPhD2003,RichterPhD2005}. It follows from Assumptions~\textit{a)} and \textit{b)} that the covariance matrix of $\V{v}$ factorizes as
\vspace*{-1mm}
\begin{align} 
 \V{Q}_{\V{v}} = 
 P\, \V{Q}_{\mathrm{s}} \otimes \V{Q}_{\mathrm{f}}
 \label{eq:C_DMC_NBFFPW}\\[-7mm]\nn
\end{align}
with $\otimes$ denoting the Kronecker product \cite{WeichselbergerPhD2003,RichterPhD2005}. The first factor is the spatial correlation matrix
\vspace*{-1mm}
\begin{align}
 \V{Q}_{\mathrm{s}}
 &= 
 \int p(\varphi)\V{s}_{\mathrm{s}}(\varphi)\V{s}_{\mathrm{s}}\herm(\varphi)\mathrm{d}\varphi
\label{eq:C_DMC_p}\\[-7mm]\nn
\end{align}
with $\V{s}_{\mathrm{s}}(\varphi) = [\text{e}^{-j2\pi f_c g(\varphi,\V{p}_{1})} \iist \cdots \iist \text{e}^{-j2\pi f_c g(\varphi,\V{p}_m)} ]\trans \in \mathbb{C}^{M\times1}$ being the array response. The second factor is the delay correlation matrix
\vspace*{-1mm}
\begin{align}
 \V{Q}_{\mathrm{f}} = \int p(\tau) 
 \V{s}_\mathrm{f}(\tau)
 \V{s}_\mathrm{f}\herm(\tau)
  \mathrm{d}\tau
 \label{C_DMC_tau}\\[-7mm]\nn
\end{align}
with
$\V{s}_\mathrm{f}(\tau)= \rmv \big[\underline{S}\big(n\Delta\big)\text{e}^{j2\pi n\Delta \tau}: n=-(N-1)/2,\iist \ldots\iist ,(N-1)/2\big]\trans\in \mathbb{C}^{N\times 1}$.

\textit{c)}\ist\ist\ist\ist 
Experimental evidence shows that the \ac{dps} typically exhibits an exponentially decaying tail \cite{RichterPhD2005, KaredalTWC2007, Andersen2007} and a smooth onset \cite{KaredalTWC2007, PedersenTAP2019}. This behaviour is well represented  by a truncated and normalized gamma PDF given by
\vspace*{-1mm}
\begin{align}
 \hspace*{-2.1mm}p(\tau) &\rmv\rmv=\rmv\rmv p(\tau;\V{\vartheta}) \nn\\
 &\rmv\rmv=\rmv\rmv\left\{\rmv\rmv\rmv
 \begin{array}{lcl}
 \frac{a}{\theta^\xi \Gamma(\xi)}
 \big(
 \tau\rmv\rmv-\rmv\rmv\beta
  \big)^{\xi-1} 
 \mathrm{e}^{
 -\frac{\tau-\beta}{\theta}
 }
 u(\tau\rmv\rmv-\rmv\rmv\beta) \rrmv\rrmv\rmv&,\rmv\rmv\rmv\rmv&\tau\in[0,T) \\
 0 \rrmv\rrmv\rmv&,\rmv\rmv\rmv\rmv&\text{elsewhere}
 \end{array}
 \right.\rmv\rmv\rmv
  \ist \label{eq:dps_parametric}\\[-7mm]\nn
\end{align}
where $u(\tau)$ is the unit step function, $\Gamma(\cdot)$ is the gamma function, and $\V{\vartheta}=[\beta\iist\iist\theta\iist\iist\xi]$ collects the onset ($\beta\rmv>\rmv0$), scale  ($\theta\rmv>\rmv0$), and shape ($\xi\rmv>\rmv0$) parameters. The normalization constant $a\rmv\rmv>\rmv\rmv0$ guarantees that $\int\rmv p(\tau;\V{\vartheta})\ist\mathrm{d}\tau\rmv\rmv=\rmv\rmv1$. The range of the parameter $\beta$ is restricted in such a way to ensure that the integral of the truncated tail of the gamma PDF is negligibly small, i.e., $a(\beta)\approx 1$ for any such values of $\beta$.

\textit{d)}\ist\ist\ist\ist We neglect the spatial correlation across antenna elements, i.e., we set $\V{Q}_{\mathrm{s}} \rmv\rmv=\rmv\rmv \V{I}_{M}$ \cite{RichterPhD2005,SalmiTSP2009}. 
This choice provides a good approximation of $\V{Q}_{\mathrm{s}}$ under the assumption of uniform \ac{aps} for the antenna-element spacings used in practice.\footnote{This assumption is exact for a uniform linear array with half-a-wavelengh element-spacing in the case of $3$~dimensional propagation with uniform direction dispersion. For horizontal-only propagation with uniform \ac{aps}, an antenna spacing equal to approximately $40$\% of the wavelength leads to practically uncorrelated entries in $\V{v}$.}

By combining Assumptions \textit{a)}\---\textit{d)} the covariance matrix $\V{Q}$ takes the following form:
\vspace*{-1.5mm}
\begin{align}
\V{Q}=\V{Q}(\V{\eta}) =  \V{I}_{M} \otimes P\ist\V{Q}_{\mathrm{f}} + \sigma^2\V{I}_{MN}
\label{eq:Cov_approx}\\[-6.5mm]\nn
\end{align}
where $\V{\eta} = [\sigma^2\iist\iist P\iist\iist \V{\vartheta}]$ with $\V{\vartheta}$ defined above and $\V{Q}_{\mathrm{f}}$
given in \eqref{C_DMC_tau} with $p(\tau)=p(\tau;\V{\vartheta})$ according to \eqref{eq:dps_parametric}. Hence, because of Assumption\ \textit{c)}, $\V{Q}(\V{\eta})$ is block-diagonal with $M$ identical $N\times N$ diagonal submatrices equal to
\vspace*{-1mm}
\begin{align}\label{eq:BDEC}
\tilde{\V{Q}}=\tilde{\V{Q}}(\V{\eta})=P\ist\V{Q}_{\mathrm{f}}
 + \sigma^2\V{I}_{N}\ist.\\[-7mm]\nn
\end{align}
	
	\section{Sparse Bayesian Formulation}\label{sec:problemformulation}
	If the number of components $K$ of the model \eqref{eq:stack_recvgen} were known, the vectors of dispersion parameters $\tilde{\V{\psi}}$ and complex amplitudes $\tilde{\V{\alpha}}$ of the \acp{sc} and the parameter vector $\V{\eta}$ of colored noise could be inferred using a standard \ac{map} or \ac{ml} estimation technique. Since we can view the family $\{\V{S}(\tilde{\V{\psi}})\}_{\tilde{\V{\psi}}\in\Psi}$ as a continuous dictionary, atomic-norm methods seem at first glance to be an inference method particularly tailored to our model. However, as detailed in the discussion of Section~\ref{sec:SBVI_algorithm}, some specificities of the model prevent a direct application of these methods.

We propose an approach inspired from \ac{sbl} \cite{TippingJMLR2001} for \ac{ssr} to include the estimation of the unknown $K$. This approach requires a two-fold modification of the generative signal model that we address below.

\subsection{Discrete-Frequency Signal Model for Inference}\label{sec:infsigmod}

In a first step the initial generative signal model \eqref{eq:stack_recvgen} is modified as follows. The number of hypothetical \acp{sc} is set to a fixed number, say $L$. Parameter $L$ is a design parameter that is selected large enough so that $K\leq L$. In addition, $L\ll NM$. Actually we only need that $L\leq MN$. The further restriction $\ll$ is for feasibility issues. Similarly as in Subsection~\ref{sec:gensigmod}, we define the vector $\V{\psi} = [\V{\psi}_1\iist\cdots\iist \V{\psi}_L]$ with entries $\V{\psi}_l = [\tau_l \iist\iist \varphi_l] \in \Psi$,
$l \in \{1,\dots,L\} \triangleq \mathcal{L}$. 
With these modifications, we arrive at the discrete-frequency signal model given by
\vspace*{-2mm}
\begin{align}\label{eq:stack_recv}
\V{y} = \V{S}(\V{\psi})\V{\alpha} + \V{n} \quad \in \mathbb{C}^{NM \times 1}\\[-7mm]\nn
\end{align}
where $\V{\alpha} = [\alpha_1 \iist\cdots\iist \alpha_{L}]\trans \in \mathbb{C}^{L\times 1}$ and $\V{S}(\V{\psi})=[\V{s}(\V{\psi}_1) \iist\cdots\iist \V{s}(\V{\psi}_{L})] \in \mathbb{C}^{NM\times L}$
with $\V{s}(\V{\psi}_l)$ defined similarly to \eqref{eq:vs}. Under the assumptions made in Subsection~\ref{sec:signalmodel}, the likelihood function of this model reads 
\begin{align}\label{eq:likelihood}
 f(\V{y}|\V{\psi},\V{\eta},\V{\alpha}) &= [ \pi^{N M} \det(\V{Q}(\V{\eta}))]^{-1}\nn\\
 &\hspace*{5mm}\times\text{e}^{-(\V{y}-\V{S}(\V{\psi})\V{\alpha})^\text{H}\V{Q}(\V{\eta})^{-1}(\V{y}-\V{S}(\V{\psi})\V{\alpha})}\\[-7mm]\nn
\ist
\end{align}
with $\det(\cdot)$ denoting the determinant of a matrix. The second step consists in specifying a hierarchical prior for each entry $\alpha_l$ in form of a Gaussian scale mixture. 
Specifically, we define
\vspace*{-4mm}
\begin{align} 
 f(\V{\alpha},\V{\gamma}) = f(\V{\alpha}|\V{\gamma}) f(\V{\gamma}) 
 =\Pi_{l \in \mathcal{L}} f(\alpha_l |\gamma_l)f(\gamma_l)
 \label{eq:priorDef}\\[-6mm]\nn
\end{align} 
where $f(\alpha_l | \gamma_l)=\sqrt{\gamma_l/2\pi}\exp\{-\gamma_l \vert\alpha_l \vert^2/2\}$, $l\in\mathcal{L}$
and $\V{\gamma} = [\gamma_1 \iist\cdots\iist \gamma_{L}]\trans \in \mathbb{R}_+^{L\times 1}$ with  $\mathbb{R}_+=\{r\in\mathbb{R}; r\geq 0 \}$. Note that all entries in $\V{\gamma}$ have the same prior with \ac{pdf} 
 $f(\gamma)$. 
These entries and their prior are referred to as hyperparameters and hyperprior, respectively. 
We postulate priors for the parameter vectors $\V{\psi}$ and $\V{\eta}$ with respective \acp{pdf} $f(\V{\psi})$ and $f(\V{\eta})$. With these specifications, the probabilistic model for inference reads
\vspace*{-4mm}
\begin{align}\label{eq:posteriorSparse} 
f(\V{\psi},\V{\eta},\V{\alpha},\V{y}) = f(\V{y}|\V{\psi},\V{\eta},\V{\alpha}) 
 f(\V{\alpha}|\V{\gamma}) f(\V{\gamma})f(\V{\psi}) f(\V{\eta})\ist.\\[-7mm]\nn
\end{align}

\subsection{Inference Method}\label{sec:sel-inf-method}

The proposed method is inspired from SBL \cite{TippingJMLR2001}. 
First it computes a \ac{map} estimate of $\V{\psi}$, $\V{\eta}$, and $\V{\gamma}$ from the joint posterior of these random vectors; then it uses this estimates to infer an approximation of the posterior distribution of $\V{\alpha}$.

The posterior \ac{pdf} $f(\V{\psi}$, $\V{\eta}$, $\V{\gamma}|\V{y})$ is obtained from \eqref{eq:posteriorSparse} by marginalizing out the complex amplitude vector $\V{\alpha}$, i.e.,
\vspace*{-1mm}
\begin{align}
f(\V{\psi},\V{\eta},\V{\gamma}|\V{y}) &\propto 
\int f(\V{y},\V{\psi},\V{\eta},\V{\alpha}) 
\text{d}\V{\alpha}\nn\\ 
&= f(\V{y}|\V{\psi},\V{\eta},\V{\gamma}) f(\V{\gamma})f(\V{\psi}) f(\V{\eta}) \label{eq:marginalPost}
 \\[-6mm]\nn
\end{align}
where
\vspace*{-1mm}
\begin{align}
 f(\V{y}|\V{\psi},\V{\eta},\V{\gamma})
&\propto \det(\M{C}(\V{\psi},\V{\eta},\V{\gamma}))^{-1}\text{e}^{-\V{y}^\text{H} \M{C}(\V{\psi},\V{\eta},\V{\gamma})^{-1} \V{y}}
\label{eq:marginalLLF}\\[-6mm]\nn
\end{align}
with $\M{C}(\V{\psi},\V{\eta},\V{\gamma}) = \V{Q}({\V{\eta}}) + \V{S}(\V{\psi}) \V{\Gamma}^{-1}\V{S}(\V{\psi})^{\text{H}}$ 
and $\V{\Gamma} = \mathrm{diag}([\gamma_1 \iist\cdots\iist \gamma_{L}])$. The \ac{map} estimates of $\V{\psi}$, $\V{\eta}$ and $\V{\gamma}$ are then computed using \eqref{eq:marginalPost}. 
From \eqref{eq:posteriorSparse} we get
\vspace*{-1mm}
\begin{align}\label{eq:post_compAmpl}
f(\V{\alpha}|\V{y},\V{\psi},\V{\eta},\V{\gamma})\propto
f(\V{y}|\V{\psi},\V{\eta},\V{\alpha}) 
 f(\V{\alpha}|\V{\gamma}),\\[-6mm]\nn
\end{align}
which is readily shown to be Gaussian with mean
\vspace*{-1mm}
\begin{align}
 \V{\mu} &= \V{\Sigma}\ist \V{S}(\V{\psi})\herm \V{Q}(\V{\eta})^{-1} \V{y} \label{eq:MMAlpha}
 \\[-6mm]\nn
\end{align}
and covariance matrix
\vspace*{-1mm}
\begin{align}
 \V{\Sigma} &= \big(\V{S}(\V{\psi})^\text{H} \V{Q}(\V{\eta})^{-1}\V{S}(\V{\psi}) + \V{\Gamma}\big)^{-1}\ist. \label{eq:VVAlpha}\\[-6mm]\nn
 \end{align} 
The approximate posterior \ac{pdf} of $\V{\alpha}$ results by plugging the \ac{map} estimates of $\V{\psi}$, $\V{\eta}$ and $\V{\gamma}$ in \eqref{eq:post_compAmpl}, and thus in \eqref{eq:MMAlpha} and \eqref{eq:VVAlpha}.

In our design, we select non-informative improper priors for $\V{\psi}$, $\V{\gamma}$ and $\V{\eta}$: $f(\V{\psi})\propto 1$, $f(\V{\gamma})\propto 1$, $f(\V{\eta})\propto 1$. With this selection, the above \ac{map} estimates coincide with the \ac{ml} estimates
\vspace*{-1mm}
\begin{align}
 (\hat{\V{\psi}}_\text{ML},\hat{\V{\eta}}_\text{ML},\hat{\V{\gamma}}_\text{ML}) &=
\argmax{\V{\psi},\V{\eta},\V{\gamma}}\ist f(\V{y}|\V{\psi},\V{\eta},\V{\gamma}) \nn\\
 &=\argmin{\V{\psi},\V{\eta},\V{\gamma}} \{
 \log(\det(\M{C}(\V{\psi},\V{\eta},\V{\gamma}))) \nn\\
 &\hspace*{10mm}+ \V{y}^\text{H} \M{C}(\V{\psi},\V{\eta},\V{\gamma})^{-1} \V{y}\}
\label{eq:T2Estimator}\\[-6mm]\nn
\end{align}
and the posterior \ac{pdf} of $\V{\alpha}$ is inferred using the approximation $f(\V{\alpha}|\V{y},\hat{\V{\psi}}_\text{ML},\hat{\V{\eta}}_\text{ML},\hat{\V{\gamma}}_\text{ML})$.

	\section{Iterative Design of the Estimator}\label{sec:SBVI_algorithm}
	Since the \ac{ml} estimator in \eqref{eq:T2Estimator} cannot be calculated analytically, even though the likelihood function is given in an analytical form, and a direct numerical solution is computationally prohibitive, we resort to a sequential update of the parameter vectors $\V{\psi}$, $\V{\eta}$, and $\V{\gamma}$ resulting in the  estimates $\hat{\V{\psi}}$, $\hat{\V{\eta}}$, and $\hat{\V{\gamma}}$. 

\subsubsection{Estimation of the Supports of the Spectral Lines}\label{sec:infdisparam}

Inserting the current estimates $\hat{\V{\eta}}$, and $\hat{\V{\gamma}}$ in \eqref{eq:T2Estimator} the new estimate of $\V{\psi}$ is computed to be
\begin{align}
\vspace*{-1mm}
 \hspace*{-2.5mm}\hat{\V{\psi}} \rmv\rmv=\rmv\rmv \argmin{\V{\psi}} \{
 \log(\det(\M{C}(\V{\psi},\hat{\V{\eta}},\hat{\V{\gamma}}))) \rmv\rmv+\rmv\rmv\V{y}^\text{H} \M{C}(\V{\psi},\hat{\V{\eta}},\hat{\V{\gamma}})^{-1} \V{y}\}.\rrmv\rrmv
 \label{eq:qpointdispersion}
  \\[-6mm]\nn
\end{align}

\subsubsection{Estimation of the Parameters of Colored Noise}\label{sec:infnoiseparam}

Similarly, the new estimate of $\V{\eta}$ is computed based on the current estimates $\hat{\V{\psi}}$, and $\hat{\V{\gamma}}$ to be
\vspace*{-1mm}
\begin{align}
\hspace*{-2.5mm}\hat{\V{\eta}} \rmv\rmv=\rmv\rmv \argmin{\V{\eta}} \{
 \log(\det(\M{C}(\hat{\V{\psi}},\V{\eta},\hat{\V{\gamma}})))\rmv\rmv +\rmv\rmv \V{y}^\text{H} \M{C}(\hat{\V{\psi}},\V{\eta},\hat{\V{\gamma}})^{-1} \V{y}\}.\rrmv\rrmv
 \label{eq:qpointnoise}
 \\[-6mm]\nn
\end{align}

\subsubsection{Estimation of the Hyperparameters}\label{sec:infgamma}

Finally, given the current estimates $\hat{\V{\psi}}$ and $\hat{\V{\eta}}$, the new estimate of $\V{\gamma}$ is updated according to
\vspace*{-1mm}
\begin{align}
\hspace*{-2.5mm}\hat{\V{\gamma}}\rmv\rmv =\rmv\rmv \argmin{\V{\gamma}} \{
 \log(\det(\M{C}(\hat{\V{\psi}},\hat{\V{\eta}},\V{\gamma})))\rmv\rmv +\rmv\rmv\V{y}^\text{H} \M{C}(\hat{\V{\psi}},\hat{\V{\eta}},\V{\gamma})^{-1} \V{y}\}.\rrmv\rrmv
 \label{eq:qpointgamma}
  \\[-6mm]\nn
\end{align}
In the sequel we consider instead of \eqref{eq:qpointgamma} a sequential method in which the estimate of each entry in $\V{\gamma}$ is updated while the estimate of the other entries are kept fixed \cite{Tipping2003}:
\vspace*{-1mm}
\begin{align}\label{eq:hyperprior}
 \hat{\gamma}_l = \left\{ \begin{array}{ll}
 (|\rho_l|^2-\zeta_l)^{-1} &, \ist \frac{|\rho_l|^2}{\zeta_l} > \kappa \\
  \infty &, \ist \frac{|\rho_l|^2}{\zeta_l}\leq \kappa \ist
  \end{array}\right. \quad l\in\mathcal{L}
    \\[-7mm]\nn
\end{align} 
with $\kappa = 1$.
In this expression
\vspace*{-1mm}
\begin{align}
\label{eq:res_var}
 \zeta_l &= \Big(\V{s}(\hat{\V{\psi}}_l)^\text{H}\V{Q}(\hat{\V{\eta}})^{-1}\V{s}(\hat{\V{\psi}}_l)- \V{s}(\hat{\V{\psi}}_l)^\text{H}\V{Q}(\hat{\V{\eta}})^{-1} \nn\\
 &\hspace*{20mm}\times\V{S}(\hat{\V{\psi}}_{\bar{l}}) \ist\hat{\V{\Sigma}}_{\bar{l}}\ist  \V{S}(\hat{\V{\psi}}_{\bar{l}})^\text{H}\V{Q}(\hat{\V{\eta}})^{-1}\V{s}(\hat{\V{\psi}}_l)\Big)^{-1} \\
 \rho_l &= \zeta_l \V{s}(\hat{\V{\psi}}_l)^\text{H}\V{Q}(\hat{\V{\eta}})^{-1}\V{y} \nn\\
 &\hspace*{5mm}- \zeta_l \V{s}(\hat{\V{\psi}}_l)^\text{H}\V{Q}(\hat{\V{\eta}})^{-1} \V{S}(\hat{\V{\psi}}_{\bar{l}}) \ist\hat{\V{\Sigma}}_{\bar{l}}\ist  \V{S}(\hat{\V{\psi}}_{\bar{l}})^\text{H}\V{Q}(\hat{\V{\eta}})^{-1}\V{y} \nn \\
  &= \zeta_l \V{s}(\hat{\V{\psi}}_l)^\text{H}\V{Q}(\hat{\V{\eta}})^{-1} \bar{\V{y}}_l
  \label{eq:res_mean}\\[-7mm]\nn
\end{align}
with 
\vspace*{-2mm}
\begin{align}
\bar{\V{y}}_l &= \V{y} - \V{S}(\hat{\V{\psi}}_{\bar{l}})\ist \hat{\V{\mu}}_{\bar{l}}
\label{eq:residual} \\
\hat{\V{\mu}}_{\bar{l}} &=\hat{\V{\Sigma}}_{\bar{l}}\ist \V{S}(\hat{\V{\psi}}_{\bar{l}})^H \V{Q}(\hat{\V{\eta}})^{-1} \V{y} \\
\hat{\V{\Sigma}}_{\bar{l}} &= (\V{S}(\hat{\V{\psi}}_{\bar{l}})^H \V{Q}(\hat{\V{\eta}})^{-1}\V{S}(\hat{\V{\psi}}_{\bar{l}}) + \hat{\V{\Gamma}}_{\bar{l}} ))^{-1} \\
\hat{\V{\Gamma}}_{\bar{l}} &=\mathrm{diag}([\hat{\gamma}_1\iist\cdots\iist \hat{\gamma}_{l-1}\iist\iist \hat{\gamma}_{l+1} \iist\cdots\iist\hat{\gamma}_{L}])
\\[-7mm] \nn
\end{align}
where $\mathrm{diag}(\cdot)$ describes a square diagonal matrix with the elements of the vector given as an argument on the main diagonal and $\hat{\V{\psi}}_{\bar{l}}=[\hat{\V{\psi}}_1 \iist \cdots\iist \hat{\V{\psi}}_{l-1}\iist\iist\hat{\V{\psi}}_{l+1}\iist\cdots\iist\hat{\V{\psi}}_{L}]$.

Note that the computation step of  $\hat{\gamma}_l$ \eqref{eq:hyperprior} contains a condition that determines when the $l$th spectral line shall be discarded ($\hat{\gamma}_l=\infty$).

\subsubsection{Estimation of the Weights}\label{eq:MAlpha}\label{eq:VAlpha}

Inserting the estimates $\hat{\V{\psi}}$, $\hat{\V{\eta}}$, and $\hat{\V{\gamma}}$ in \eqref{eq:post_compAmpl} yields the Gaussian \ac{pdf} 
with mean (see \eqref{eq:MMAlpha})
\vspace*{-1mm}
\begin{align}
\hat{\V{\mu}} &= \hat{\V{\Sigma}}\ist \V{S}(\hat{\V{\psi}})^\text{H} \V{Q}(\hat{\V{\eta}})^{-1} \V{y} \label{eq:HatMAlpha}\\[-6mm] \nn
\end{align}
and covariance matrix (see \eqref{eq:VVAlpha})
\vspace*{-1mm}
\begin{align}
\hat{\V{\Sigma}} &= \big(\V{S}(\hat{\V{\psi}})^\text{H} \V{Q}(\hat{\V{\eta}})^{-1}\V{S}(\hat{\V{\psi}}) + \hat{\V{\Gamma}}\big)^{-1}\ist
\label{eq:HatVAlpha}\\[-6mm] \nn
\end{align} 
that is used as an approximation of the posterior \ac{pdf} of $\V{\alpha}$. In \eqref{eq:HatVAlpha}, $\hat{\V{\Gamma}} = \mathrm{diag}([\hat{\gamma}_1 \iist \cdots \iist \hat{\gamma}_L])$.

\subsubsection{Fitting of the Pruning Threshold $\kappa$}

Numerical experiments have shown that the iterative algorithm obtained in the above subsections overestimates the number of spectral lines and thereby returns estimates of spurious components. This bias in the number of detected components increases when either the \ac{snr} or the number of samples increases \cite[Subsec.\ V.A]{ShutinTSP2011}, \cite{ShutinArxiv2015}. Following the approach adopted in \cite{NadlerTSP2011,TalebiTSP2015}, we increase the initial threshold $\kappa=1$ in the pruning condition \eqref{eq:hyperprior} to $\kappa=\kappa^\ast>1$. The value $\kappa^\ast$ is set in such a way to reduce the bias. The next section describes in detail this procedure, which yields $\kappa^\ast$ given in \eqref{eq:kappafinalval}.	

\subsection*{Discussion}\label{s:Discussion}

The updating step \eqref{eq:qpointdispersion} in its form looks very similar to the classical unconstrained (also called stochastic) \ac{ml} estimator in sensor array signal processing \cite{Ottersten1993} with the additional assumption that the precision matrix of the weights be diagonal, i.e. equal to $\V{\Gamma}$ as a result of the gamma-Gaussian hierarchical model.\footnote{Sensor array signal processing considers a signal model similar to \eqref{eq:stack_recv} where the entries of $\V{y}$ are the outputs of an array of sensors, $\V{S}(\V{\psi})$ is the array response matrix, $\V{\psi}$ and $\V{\alpha}$ contain respectively the dispersion parameters and the amplitudes of the sources, and $\V{n}$ is the measurement noise vector \cite{Ottersten1993}. The number of sources is assumed to be known and smaller than the number of sensors in order for the model parameters to be identifiable. In practice the number of sources is estimated using an additional model-order selection procedure based on an information theoretic criterion, see Section~\ref{sec:Introduction}.} Despite the resemblance \eqref{eq:qpointdispersion} is not an instance of unconstrained \ac{ml} estimation. Unconstrained \ac{ml} estimation requires a scenario where at least as many observations (snapshots, assumed uncorrelated) as the number of sensors are collected\footnote{This condition ensures that the sample array covariance matrix has full rank, which is a mandatory condition in the derivation of the unconstrained \ac{ml} estimator.}, while in our scenario only one observation, i.e. $\V{y}$ in \eqref{eq:stack_recv}, is available. Our estimator also deviates from being an instance of \ac{sbl} \cite{TippingJMLR2001} in three respects: (a) the underlying model of \ac{sbl} is undetermined, which is not the case for our model \eqref{eq:stack_recv} with $L\rmv\rmv\ll\rmv\rmv NM$; (b) the ``dictionary matrix'', namely $\V{S}(\V{\psi})$ in \eqref{eq:stack_recv}, is not fixed but is parameterized by the continuous parameter vector $\V{\psi}$ that is estimated; and (c) the inherent threshold of \ac{sbl} is adapted to control the probability of detecting spurious \acp{sc}.\footnote{Strictly speaking, \ac{sbl} is derived under the assumption of \ac{awgn}. It can be straightforwardly applied when noise is non-white, by merely applying a whitening filter first.} Our method belongs to the parametric class estimators in the nomenclature introduced in \cite{Stoica2011}.

\Acf{atm} provides an elegant, natural means to operate with a continuous, i.e. infinite, dictionary in \ac{ssr} \cite{BhaskarTSP2013,YanXie:TSP2015}.
At first glance this method looks promising for dealing with the continuous dictionary $\{\V{S}(\V{\psi})\}_{\V{\psi}\in\Psi}$ in our problem at hand. However, some specificities of the generic model \eqref{eq:stack_recvgen} prevent its straightforward application to our scenario. Note that \ac{atm} primarily ``denoises'' the observed signal with the estimation of the spectral lines being subsequently performed based on this denoised signal. While the estimation problem can be solved with an exact semi-definite program when the dispersion domain is one-dimensional \cite{BhaskarTSP2013}, only an approximate such program could be formulated to date for higher dimensional dispersion domains \cite{ChiTSP2015}.\footnote{In \cite{ChiTSP2015} the matrix-enhancement-matrix-pencil method \cite{HuaTSP1992} is used to compute estimates of the support of spectral lines from the denoised signal.} In addition, \ac{atm} operates on Nyquist-sampled signals and requires knowledge of the noise characteristics, e.g. its spectral height when noise is white. These conditions do not hold in our application scenario.

	\section{Computation of the Pruning Threshold}\label{sec:analyPrun}
	To compute the threshold value $\kappa^*$ we adapt the approach described in \cite{ShutinArxiv2015} to our application scenario; see also \cite{NadlerTSP2011} for a similar approach applied to constrained ML estimation. To make it tractable the analysis is carried out under the following assumptions. 
\begin{assumption}\label{as:KnoCovM}
The spatial and frequency apertures \cite{Johnson1993} of the sounding equipment are centro-symmetric\footnote{Specifically, referring to Subsec.~\ref{sec:rsigmod} for any $m \in \mathcal{M}$, there exists an index $m' \in\mathcal{M}$ such that $\V{p}_{m'} - \V{p}=-(\V{p}_m-\V{p})$. The statement for the vector defining the frequency aperture is similar.} \cite{Pressman1998}. 
Furthermore,  
$\V{s}_\mathrm{f}=\V{s}_\mathrm{f}(0)$, 
see text below \cref{C_DMC_tau}, fulfils $\V{J}\V{s}_\mathrm{f}=\V{s}_\mathrm{f}^*$, where $\V{J}$ is the exchange or reversal matrix \cite[Sec.~4.8]{stoica2005spectral}. The covariance matrix $\V{Q}$ in \cref{eq:Cov_approx} is known. 
\end{assumption}

It is shown in \cite{LeitingerAsilomar2020} that as a result of the first part in the assumption the matrix $\V{Q}_{\V{v}}$ in \eqref{eq:C_DMC_NBFFPW} is centro-hermitian\footnote{Since these matrices are hermitian, their centro-hermitian property implies per-symmetry.} \cite{Pressman1998} and therefore $\V{Q}$ in \eqref{eq:Cov_approx} too. 
The next assumption reflects an empirical evidence based on extensive simulations of the proposed algorithm. 

\begin{assumption}\label{as:Regime}
Asymptotically as the dimension $MN$ grows large the estimator in Section~\ref{sec:SBVI_algorithm} with $\kappa=1$ exhibits the following behaviour: (a) it resolves all $K$ active \acp{sc} and accurately estimates their parameters, i.e. without loss of generality $\hat{\V{\psi}}_l\approx\tilde{\V{\psi}}_l$ for $l=1,\ldots,K$; (b) it computes estimates $\hat{\V{\psi}}_l$, $l=K+1,\ldots,L$ of $L-K$ (spurious) \ac{sc} components in such a way that with high probability $\V{s}(\hat{\V{\psi}}_l)$ is nearly orthogonal to any columns of $\V{S}(\hat{\V{\psi}}_{\bar{l}})$ for each $l=K+1,\ldots,L$. 
\end{assumption}
As a result of Assumption~\ref{as:Regime}, as $MN$ grows large, with high probability \eqref{eq:res_var}  and \eqref{eq:res_mean} can be approximated for $l=K+1,\ldots,L$ as $\zeta_l \approx \bar{\zeta}(\hat{\V{\psi}}_l)$ and $\rho_l \approx \bar{\rho}(\hat{\V{\psi}}_l)$, respectively, where we have defined $\bar{\zeta}(\V{\psi}_\cdot)= (\V{s}(\V{\psi}_\cdot)^\text{H}\V{Q}^{-1} \V{s}(\V{\psi}_\cdot))^{-1}$ and $\bar{\rho}(\V{\psi}_\cdot)= \bar{\zeta}(\V{\psi}_\cdot) \V{s}(\V{\psi}_\cdot)^\text{H}\V{Q}^{-1} \V{n}$, respectively, with $\V{\psi}_\cdot=[\tau \iist\iist \varphi] \in\Psi$. Therefore, the probability that the algorithm decides that the $l$th component ($l=K+1,\ldots,L$) is active, i.e. $\hat{\gamma}_l<\infty$ in \eqref{eq:hyperprior}, with threshold set to $\kappa$ is close to 
\vspace*{-1mm}
\begin{align}
P_\text{f}(\kappa) =\mathbb{P} \Big[ \sup_{{\V{\psi}_\cdot} \in \Psi}
\vert \bar{\rho}({\V{\psi}_\cdot})\vert^2/\bar{\zeta}(\V{\psi}_\cdot) 
\geq \kappa \Big]
\label{eq:pf}
\\[-6mm]\nn
\end{align}
when $MN$ is sufficiently large. Let us consider the circularly-symmetric complex Gaussian random field on $\Psi$ defined as
\vspace*{-2mm}
\begin{align}\label{eq:px}
x(\V{\psi}_\cdot) \rmv=\rmv\frac{\bar{\rho}({\V{\psi}_\cdot})}{\bar{\zeta}(\V{\psi}_\cdot)^{1/2}} \rmv=\rmv
\frac{\V{s}(\V{\psi}_\cdot)^\text{H}\V{Q}(\V{\eta})^{-1}\V{n}}
{[\V{s}(\V{\psi}_\cdot)^\text{H}\V{Q}(\V{\eta})^{-1}\V{s}(\V{\psi}_\cdot)]^{1/2}}  
\ist.\\[-5mm]\nn
\end{align} 
with $\quad\V{\psi}_\cdot\in\Psi$. Then, \eqref{eq:pf} can be recast as 
\vspace*{-1mm}
\begin{align}\label{eq:PA}
P_\text{f}(\kappa) = \mathbb{P} \Big[ \sup_{{\V{\psi}_\cdot} \in \Psi} \vert x({\V{\psi}_\cdot})\vert^2  \geq \kappa \Big]\ist.
\\[-6mm]\nn
\end{align}

\begin{theorem}
Under Assumption~\ref{as:KnoCovM} 
we have the asymptotic equivalence
\vspace*{-1mm}
\begin{align}\label{eq:PAtheorem1}
P_\mathrm{f}(\kappa)
&\sim  
\left[\frac{1}{\pi}\int_{\Psi}\sqrt{\det(\M{\Lambda}({\V{\psi}_\cdot}))}\iist\mathrm{d}{\V{\psi}_\cdot}\right]\ist \kappa \ist\text{e}^{-\kappa}
\,\, ,
 \,\,\kappa \rightarrow \infty\ist.\\[-6mm]\nn
\end{align}
Furthermore,
\vspace*{-1mm}
\begin{align}
&\frac{1}{\pi}\int_{\Psi}\rmv\rmv\sqrt{\det(\M{\Lambda}({\V{\psi}_\cdot}))}\ist\mathrm{d}{\V{\psi}_\cdot}\nn\\
&\hspace*{0mm}=\rmv\rmv4\pi\rmv\rmv\int_{0}^{\Delta^{-1}}\hspace*{-3mm}\int_{0}^{2\pi}\hspace*{-1mm}
\Big[\Big(\frac{1}{M}\hspace*{-1mm}\sum\limits_{m \in \mathcal{M}}\hspace*{-1mm}d_m^2({\varphi})\Big)a(\tau)
\Big]^{1/2}\hspace*{-1mm}f_\text{c} 
\,b(\tau)  \mathrm{d}{\tau}\mathrm{d}{\varphi}\ist. \label{eq:SDM}
\\[-6mm]\nn
\end{align}
Here, $\M{\Lambda}({\V{\psi}_\cdot})$ is the non-negative definite matrix given in \eqref{eq:cov_randomfield},
$d_m({\varphi}) = \partial g(\varphi, \V{p}_m))/\partial\varphi$, $m \in \mathcal{M}$, $b(\tau)=\big[\big(\dot{\underline{\V{s}}}(\tau)^\text{H}\tilde{\M{Q}}^{-1}\dot{\underline{\V{s}}}(\tau)\big)/\big(4\pi^2 \underline{\V{s}}(\tau)^\text{H}\tilde{\M{Q}}^{-1} \underline{\V{s}}(\tau)\big)\big]^{1/2}$, 
$ a(\tau) \rmv\rmv=\rmv\rmv 1  - \Re\big\{ \dot{\underline{\V{s}}}(\tau)^\text{H}\tilde{\M{Q}}^{-1} \underline{\V{s}}(\tau)\big\}^2 /\big(\big(\underline{\V{s}}(\tau)^\text{H}$ $\tilde{\M{Q}}^{-1} \underline{\V{s}}(\tau)\big)^2 \big(\dot{\underline{\V{s}}}(\tau)^\text{H}\tilde{\M{Q}}^{-1} \dot{\underline{\V{s}}}(\tau)\big)^2\big)$ with $\tilde{\V{Q}}$ given in \eqref{eq:BDEC} and 
$\dot{\underline{\V{s}}}(\tau) = \partial \underline{\V{s}}(\tau)/\partial\tau $.
\end{theorem}

The term $[\frac{1}{M}\sum_m d_m^2({\varphi})]^{1/2}$ incorporates the impact of the array aperture, while $b(\tau)$ and $a(\tau)$ incorporate the impact of the frequency aperture (spectrum $S(f)$) and colored noise.
\begin{proof}
As shown in \cite{LeitingerAsilomar2020} it follows from Assumption~\ref{as:KnoCovM} that the real and imaginary parts of the Gaussian field $x({\V{\psi}_\cdot})$ in \eqref{eq:px} exhibits the following properties:
\begin{enumerate}[leftmargin=6mm]
\item They have equal constant variance:
$\mathbb{E}\big[ \big|\Re\{x({\V{\psi}_\cdot})\big|^2\big] =\mathbb{E}\big[ \big|\Im\{x({\V{\psi}_\cdot})\big|^2\big] = 1/2$, $\V{\psi}_\cdot\in\Psi$. 
\item They are independent: 
$\mathbb{E}[ \Re\{x({\V{\psi}_\cdot})\}\Im\{x({\V{\psi}_\cdot}')\}] \rmv\rmv=\rmv\rmv 0$, $\V{\psi}_\cdot,\V{\psi}_\cdot' \in\Psi$.
\end{enumerate}
Since $\Re\{x({\V{\psi}_\cdot})\}$ and $\Im\{x({\V{\psi}_\cdot})\}$ are independent $2\vert x({\V{\psi}_\cdot})\vert^2$ 
is a random field $\chi^2$ on $\Psi$ with two degrees of freedom 
\cite{
WorsleyAIAP1994,Adler2015ApplicationsRandomFieldsGeometry}.\footnote{The real and imaginary parts of  $\sqrt{2}\ist x({\V{\psi}_\cdot})$ have unit variance, in accordance with the definition of a $\chi^2$ process.} 
Note that unless the \ac{dc} vanishes, i.e. $P=0$, see \eqref{eq:Cov_approx}, the Gaussian field $x({\V{\psi}_\cdot})$ is non-stationary and so is $2\vert x({\V{\psi}_\cdot})\vert^2$. 
The probability that the $\chi^2$ field exceeds a threshold is asymptotically equivalent to the probability of the field's excursion above the threshold when said threshold grows large \cite{WorsleyAIAP1994,Adler2015ApplicationsRandomFieldsGeometry}. 
Specifically, by
applying Weyl's tube formula \cite[Theorem 3.3.1]{Adler2015ApplicationsRandomFieldsGeometry} to $2\vert x({\V{\psi}_\cdot})\vert^2$ 
and making use of \cite[Theorem 4.4.1]{Adler2015ApplicationsRandomFieldsGeometry} combined with 
\cite[Section 4.5.2]{Adler2015ApplicationsRandomFieldsGeometry} we obtain 
\vspace*{-1mm}
\begin{align}\label{eq:exclusprob}
&\mathbb{P} \Big[ \sup_{{\V{\psi}_\cdot}} 2\vert x({\V{\psi}_\cdot})\vert^2
\rmv\geq\rmv 2\kappa \Big] \nn\\
&\hspace*{10mm}\sim \left[\int_{\Psi}\frac{1}{\pi}\sqrt{\det(\M{\Lambda}({\V{\psi}_\cdot}))}\iist\mathrm{d}{\V{\psi}_\cdot}\right]\ist \kappa \ist\text{e}^{-\kappa} 
\quad \kappa \rightarrow \infty\ist.
\\[-6mm]\nn
\end{align}
In this expression $\M{\Lambda}({\V{\psi}_\cdot}) \in \mathbb{R}^{2\times2}$ is the covariance matrix
\vspace*{-1mm}
\begin{align}
\M{\Lambda}({\V{\psi}_\cdot}) &=
\mathbb{E}\left[\frac{\partial x({\V{\psi}_\cdot})}{\partial {\V{\psi}_\cdot} }\left[\frac{\partial x({\V{\psi}_\cdot})}{\partial {\V{\psi}_\cdot}}\right]\herm \right] \nn\\
&=
\mtx{cc}{ \mathbb{E}\Big[\frac{\partial x({\V{\psi}_\cdot})\partial x({\V{\psi}_\cdot})^*}{\partial {\tau}^2 }\Big]  & \mathbb{E}\Big[\frac{\partial x({\V{\psi}_\cdot})\partial x({\V{\psi}_\cdot})^*}{\partial {\tau} \partial {\varphi} }\Big] \\[2mm]  \mathbb{E}\Big[\frac{\partial x({\V{\psi}_\cdot})\partial x({\V{\psi}_\cdot})^*}{\partial{\varphi} \partial {\tau} }\Big] & \mathbb{E}\Big[\frac{\partial x({\V{\psi}_\cdot})\partial x({\V{\psi}_\cdot})^*}{\partial {\varphi}^2 }\Big]}\ist.
\label{eq:cov_randomfield}
\\[-6mm]\nn
\end{align}
As shown in \cite[Appendix~A]{GreLeiFleWit:Arxiv2023} the entries in \eqref{eq:cov_randomfield} read
\begin{align}
\mathbb{E}\left[\frac{\partial x({\V{\psi}_\cdot})\partial x({\V{\psi}_\cdot})^*}{\partial {\tau}^2 }\right] &= 4\pi^2 a(\tau) b^2({\tau})
\label{eq:deriv_tau}\\[3mm]
\mathbb{E}\left[\frac{\partial x({\V{\psi}_\cdot})\partial x({\V{\psi}_\cdot})^*}{\partial {\varphi}^2 }\right] &= \frac{4\pi^2}{M} \sum_{m \in \mathcal{M}}d_m^2({\varphi}) f_\mathrm{c}^2 \\[3mm]
\mathbb{E}\left[\frac{\partial x({\V{\psi}_\cdot})\partial x({\V{\psi}_\cdot})^*}{\partial {\tau}\partial{\varphi} }\right] &= \mathbb{E}\left[\frac{\partial x({\V{\psi}_\cdot}) \partial x({\V{\psi}_\cdot})^*}{\partial {\varphi} \partial {\tau} }\right]= 0\ist. \label{eq:deriv_phi}
\end{align}
The right-hand side in \eqref{eq:SDM} follows then from \eqref{eq:deriv_tau}\--\eqref{eq:deriv_phi}.
\end{proof}

The function of $\kappa$ in the asymptotic equivalence \eqref{eq:PAtheorem1} provides a tight approximation of $P_\text{f}(\kappa)$ versus $\kappa$ for $\kappa$ sufficiently large. 
Thus, by taking the inverse of that function and evaluating it at a target probability value, say $\epsilon$, we obtain a threshold, say $\kappa^{\star}=\kappa^{\star}(\epsilon)$, that yields $P_\text{f}(\kappa^\star)$ close to $\epsilon$. 
The next lemma essentially gives this inverse function.

\begin{lemma}
Given $\epsilon\in(0,q/e]$ with $e$ denoting Euler's number and $q=\int_{\Psi}\frac{1}{\pi}\sqrt{\det(\M{\Lambda}({\V{\psi}_\cdot}))}\iist\mathrm{d}{\V{\psi}_\cdot}$ the asymptotic expression in \cref{eq:exclusprob}, and thereby in \cref{eq:PAtheorem1}, is upper-bounded by $\epsilon$ provided $\kappa$ satisfies
\vspace*{-2mm}
\begin{align}
\kappa &\geq \kappa^\star(\epsilon) = - \mathrm{W}_{-1}(-\epsilon/q)\geq 1
\label{eq:kappafinalval}\\[-6mm]\nn
\end{align}
where $\text{W}_{-1}:[-e^{-1},0)\mapsto\mathbb{R}$ is the second real branch of the Lambert-W function \cite{Corless1996}.
\end{lemma}

\begin{proof}
With the definition of $q$ the right-hand expression in \eqref{eq:exclusprob} reads $q\kappa \text{e}^{-\kappa}$. Clearly, this expression determines a non-increasing function of $\kappa$ defined on $[1,\infty)$ with range $(0,q/e]$. Given $\epsilon\in(0,q/e]$ we seek the minimum value of $\kappa\in[1,\infty)$ such that 
$q\kappa \text{e}^{-\kappa} \leq \epsilon$ holds. Obviously the sought value solves $q\kappa \text{e}^{-\kappa} = \epsilon$, i.e. equals $\kappa^\star(\epsilon)$ in \eqref{eq:kappafinalval}.
\end{proof}
We conclude from \eqref{eq:PA}, \eqref{eq:PAtheorem1} and \eqref{eq:kappafinalval} that $P_\text{f}(\kappa^\star(\epsilon))\approx \epsilon$ for $\epsilon$ sufficiently small and $MN$ sufficiently large. Thus, the function $\kappa^\star(\epsilon)$ provides a means to control the probability of detecting spurious components provided $MN$ is sufficiently large. We can use the following asymptotic behavior of the function $\mathrm{W}_{-1}(u)$ as $u\rightarrow 0$
to obtain a tight approximation of $\kappa^\star(\epsilon)$ for $\epsilon$ small:
\vspace*{-1mm}
\begin{align} 
\mathrm{W}_{-1}(u)=\log(-u) - \log(-\log(-u)) + o(1)\ist,\ist u\rightarrow 0\ist\\[-6mm]\nn
\end{align}
where $o(\cdot)$ denotes the little-o notation \cite{Corless1996}. Making use of this identity, the equality in \eqref{eq:kappafinalval} can be recast as
\vspace*{-1mm}
\begin{align}
\kappa^\star(\epsilon) = - \log(\epsilon/q) + \log(-\log(\epsilon/q)) + o(\epsilon)\ist,\ist \epsilon\rightarrow 0\ist.\\[-6mm]\nn
\end{align}
The right-hand expression with the term $o(\epsilon)$ dropped provides a tight approximation of $\kappa^\star(\epsilon)$ when $\epsilon$ is sufficiently small, provided $MN$ is sufficiently large.

\subsection{Examples:}

We illustrate the right-hand expression in \eqref{eq:SDM} with two examples. We consider a scenario with a sounding signal exhibiting a constant spectrum over its bandwidth, i.e., Assumption~\ref{as:KnoCovM} is fulfilled, and \ac{awgn} only.  	
In this case \eqref{eq:SDM} becomes
$
		 \textstyle
		 4 \pi \sqrt{\frac{N^2-1}{12}}
		\ist\ist
		\int_0^{2\pi}f_\mathrm{c}\sqrt{\textstyle\frac{1}{M}\sum_{m \in \Set{M}} d_m^2({\varphi})} \iist \mathrm{d}{\varphi}\ist
$.

\begin{proof}
In this case, $b(\tau)\rmv\rmv=\rmv\rmv \Delta\sqrt{(N^2 -1)/12}=B/N
\sqrt{(N^2 -1)/12}$ 
and $a(\tau) = 1$.
\end{proof}

\setlength{\textfloatsep}{3pt}
\begin{algorithm}[!t]
	\SetInd{0.5em}{0.5em}
	\SetAlgoLined
	\DontPrintSemicolon
	\SetKwInOut{Input}{Input}
	\SetKwInOut{Output}{Output}
	\SetKwFunction{search}{search}
	\SetKwFunction{refine}{refine}
	\SetKwFunction{prune}{prune}
	\SetKwFunction{initalizeDC}{initalizeDC}
	\Input{Measurement vector $\V{y}$}
	\Output{$\hat{L}$, $\hat{\V{\psi}}$, $\hat{\V{\eta}}$, $\hat{\V{\mu}}$, and  $\hat{\V{\Sigma}}$
	}
	$\hat{\V{\psi}}, \hat{\V{\gamma}}, \hat{\V{\mu}}, \hat{\V{\Sigma}}  \leftarrow [\iist],[\iist],[\iist],[\iist]$\;
	{ $\hat{\V{\eta}} \leftarrow [\hat{\sigma}^2\iist\iist \hat{P} \iist\iist \hat{\V{\vartheta}}]$
		\linebreak with $\hat{\sigma}^2 \leftarrow \frac{\|\V{y}\|^2}{NM}$ , $\hat{P} \leftarrow 0$, $\hat{\V{\vartheta}} \leftarrow\V{0}$ (AWGN only), and $\textrm{initFlag}\leftarrow \textrm{true}$\;}
	
	\Do{$\|\hat{\V{\mu}}\|_0$ changes or the maximum number of cycles is reached}{
		\search{}\;
		\refine{}\;
		$\hat{\V{\mu}},\hat{\V{\Sigma}}\leftarrow$ \eqref{eq:HatMAlpha} and \eqref{eq:HatVAlpha}, respectively\;
	}
	$\hat{L} \leftarrow \|\hat{\V{\mu}}\|_0$\footnotemark\;
	\caption{Main}\label{alg:sbl}
\end{algorithm}

\footnotetext{The operator $\|\cdot\|_0$ gives the number of non-zero elements of the vector given as argument.}%
\setcounter{myproc}{0}
\begin{myproc}[!t]
	\SetInd{0.5em}{0.5em}
	\SetAlgoLined
	\DontPrintSemicolon
	\SetKwFunction{search}{search}
	\SetKwFunction{refine}{refine}
	\SetKwFunction{initalizeDC}{initalizeDC}
	\SetKwFunction{prune}{prune}
	\SetKwProg{proc}{Procedure}{}{}
	\proc{\search{}}{
		\If{$\|\hat{\V{\gamma}}\|_0<L$}
		{
			\vspace*{0.5mm}
			$l \leftarrow \|\hat{\V{\gamma}}\|_0 + 1$ \;
			$ \hat{\V{\psi}}_l \leftarrow \argmax{\V{\psi}} \frac{|\rho_l|^2}{\zeta_l}$ using \eqref{eq:res_var} and \eqref{eq:res_mean}\;
			$\hat{\gamma}_l\leftarrow(|\rho_l|^2\rmv-\rmv\zeta_l)^{-1}$ \;
			
			append $\hat{\V{\psi}}_l$ to $\hat{\V{\psi}}$ and $\hat{\gamma}_l$ to $\hat{\V{\gamma}}$\;
		}
	}
	\caption{Search}\label{proc:search}
\end{myproc}

\setcounter{myproc}{1}
\begin{myproc}[!t]
	\SetAlgoLined
	\DontPrintSemicolon
	\SetKwFunction{search}{search}
	\SetKwFunction{refine}{refine}
	\SetKwFunction{initializeDC}{initializeDC}
	\SetKwFunction{prune}{prune}
	\SetKwProg{proc}{Procedure}{}{}
	\proc{\refine{}}{
		\Do{not converged}
		{
			$\hat{\V{\eta}} \leftarrow$ update according to \eqref{eq:qpointnoise}\;
			$\hat{\V{\psi}}\leftarrow$ update according to \eqref{eq:qpointdispersion}\;
			$\hat{\kappa}^\star(\epsilon) \leftarrow$ \eqref{eq:kappafinalval} \;
			$\hat{\V{\gamma}} \leftarrow$ update according to \eqref{eq:hyperprior},  $\kappa\iist=\iist\hat{\kappa}^\star(\epsilon)$\;
			\initializeDC{}\;
			\For{$l \leftarrow 1,\dots,\|\hat{\V{\gamma}}\|_0$}
			{
				\If{$\hat{\gamma_l}=\infty$}{remove $l$th component from $\hat{\V{\psi}}$ and $\hat{\V{\gamma}}$}
			}
		}
	}
	\caption{Refine}\label{proc:refine}
\end{myproc}

\setcounter{myproc}{2}
\begin{myproc}[!t]
	\SetInd{0.5em}{0.5em}
	\SetAlgoLined
	\DontPrintSemicolon
	\SetKwFunction{search}{search}
	\SetKwFunction{refine}{refine}
	\SetKwFunction{initializeDC}{initializeDC}
	\SetKwProg{proc}{Procedure}{}{}
	\proc{\initializeDC{}}{
		\If{(any entry of $\hat{\V{\gamma}}$ is $\infty$ or $L$ reached) and $\textrm{initFlag}$ is $\textrm{true}$}
		{
			$ \hat{P}\leftarrow \hat{\sigma}^2/(2\Delta)$\;
			$ \hat{\V{\vartheta}} \leftarrow [1\text{m}/c\iist\iist T/2 \iist\iist 2]$\;
			$\hat{\V{\eta}} \leftarrow [\hat{\sigma}^2/2 \iist\iist \hat{P} \iist\iist \hat{\V{\vartheta}}]$\;
			\refine{}\;
			$\textrm{initFlag}\leftarrow \textrm{false}$\;
		}
	}
	\caption{Initialize \ac{dc} parameters}\label{proc:initalizeDMC}
\end{myproc}

If furthermore the array is uniform, square, of dimensions $M'\rmv\rmv\times\rmv\rmv M'$, and with equal 
inter-element spacing $w>0$, the above expression further simplifies to $ 8 \pi^2 \textstyle\sqrt{\frac{N^2-1}{12}} \textstyle\sqrt{ f_\mathrm{c}^2 \frac{w^2}{c^2}\frac{M-1}{12}}$ with $M = M'^{\ist 2}$.
\begin{proof}
	The square aperture function of a rectangular uniform array is given as
		$\frac{1}{M}\sum\limits_{m \in \Set{M}}\rmv\rmv\rmv\rmv d_m^2({\varphi}) = 
		\textstyle M'\frac{w^2}{c^2} \sin^2({\varphi}\rmv\rmv-\rmv\rmv\psi)\frac{M'(M'^2\rmv\rmv-\rmv\rmv 1)}{12} + M'\frac{w^2}{c^2} \cos^2({\varphi}\rmv\rmv-\rmv\rmv\psi)\frac{M'(M'^2\rmv\rmv-\rmv\rmv 1)}{12}\rmv\rmv=\rmv\rmv \frac{w^2}{c^2}\frac{M\rmv\rmv-\rmv\rmv 1}{12}\ist.
	$
\end{proof}

	\vspace*{-2mm}
	\section{Implementation}\label{sec:implementation}
	\begin{figure*}[!t]
	\vspace{-4mm}
	\captionsetup[subfigure]{captionskip=0pt}
	\subfloat[\label{fig:PaPm_knownC}]{\includegraphics[width=0.32\textwidth,height=0.15\textwidth]{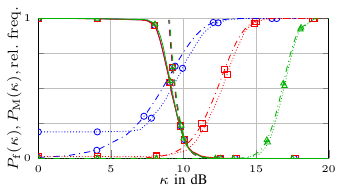}}\hspace*{-1mm}
	\subfloat[\label{fig:PaPm_unknownC}]{\includegraphics[width=0.32\textwidth,height=0.15\textwidth]{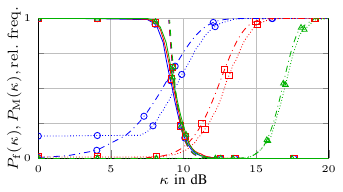}}\hspace*{-1mm}
	\subfloat[\label{fig:PaPm_UWBC}]{\includegraphics[width=0.32\textwidth,height=0.15\textwidth]{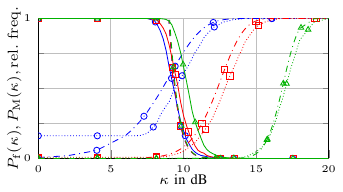}}\\[0mm]
	\centering
	\includegraphics{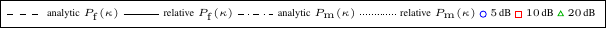}
	\vspace{-2mm}
	\caption{Comparison of the probalilities $P_\text{f}(\kappa)$ and $P_\text{m}(\kappa)$ with their respective relative frequencies computed from $1000$~trials with $\SNR$ as a parameter. The three scenarios corresponding each to specific assumptions on $\V{Q}$ are described in the text.}	
	\label{fig:PaPm}
	\vspace*{-3mm}
\end{figure*}

The pseudocode of the proposed algorithm is given in Algorithm~\ref{alg:sbl}. It has two main stages: a search and a refine procedure, described in Procedure~\ref{proc:search} and Procedure~\ref{proc:refine}, respectively. After initialization the two procedures are executed sequentially in a do-while loop until a stopping criterion is met. 
Specifically, Algorithm~\ref{alg:sbl} implements a bottom-up strategy: starting with an empty model, i.e. $\hat{L}=0$, at each iteration of the do-while-loop Procedure~\ref{proc:search} searches and adds a candidate \ac{sc} in the current pool of so far buffered candidates \acp{sc}. Procedure~\ref{proc:refine} estimates and/or re-estimates the parameters of all candidate \acp{sc} in the pool, and possibly removes candidate \acp{sc} to finally yield an updated pool of $\hat{L}$ candidate \acp{sc}. 
The algorithm terminates once the number $\hat{L}$ of \acp{sc} in the pool and their parameter estimates as well as the estimated parameters of the \ac{dc} are converged. It then returns these converged values as the model estimates.

The initial iterations in the do-while-loop of Algorithm~\ref{alg:sbl} 
are executed while considering measurement noise only, i.e. by using $\V{Q}$ in \eqref{eq:Cov_approx} with $P$ set to zero whenever $\V{Q}$ occurs in the update equations of Procedures~\ref{proc:search}~and~\ref{proc:refine}. This is carried out until a first candidate \ac{sc} in the pool is pruned in Procedure~\ref{proc:refine} or $\hat{L}$ reaches a predefined maximum number $L$, in which case Procedure~\ref{proc:initalizeDMC} is executed to initialize the parameters of the \ac{dc}. Once the initialization is completed, the noise variance estimate $\hat{\sigma}^2$ aggregates a contribution from the \ac{dc}. The total estimated power over the bandwidth computed from this value is distributed evenly between noise and the \ac{dc} in Procedure~3. This explains the factor $1/2$ in Lines 3 and 5.
From then on the estimates of the parameters of the \ac{dc} \ac{dps} are updated, i.e. the full covariance matrix $\V{Q}$ in \eqref{eq:Cov_approx} is accounted for in the update equations of Procedures~\ref{proc:search}~and~\ref{proc:refine}.

	\section{Numerical and Experimental Results}\label{sec:results}
	To validate the proposed algorithm, we first test it in Subsection~\ref{sec:SyntheticChannels} with synthetically generated measurements according to the model in \eqref{eq:stack_recvgen} with a covariance matrix $\V{Q}$ given in \eqref{eq:C_DMC_full} and \eqref{eq:Cov_approx}.
Then, in Subsection~\ref{sec:MeasData} we apply the algorithm to measurements collected in an indoor environment. 

\subsection{Synthetic Radio Channels}\label{sec:SyntheticChannels}

In this study, the signal spectrum $\underline{S}(f)$ has a root-raised cosine shape with roll-off factor  $0.6$ and bandwidth $B = 1.6$\,GHz centered at $f_\text{c} = 6$\,GHz. The 3-dB bandwidth $\tilde{B}$ is $1$\,GHz and yields the \ac{rrl} $1/\tilde{B}=1$\,ns. Each numerical investigation involves $1000$ simulation trials. In each trial the Gaussian \ac{dc} vector $\V{v}$ (see text below \eqref{eq:sm}) is generated using \eqref{eq:dps_parametric} with $\beta=1\text{m}/c$, $\theta=5$\,ns and $\xi=1.8$. The power $P$ is specified through the specular-to-dense-ratio $\SDR = 10\log_\mathrm{10}\big(\frac{1}{M} \| \sum_{k \in \mathcal{K}} \tilde{\alpha}_k \V{s}(\tilde{\V{\psi}}_k)\|^2/(PB)\big)$. In addition, the Gaussian noise vector $\V{w}$ is generated with component variance $\sigma^2$ specified through the signal-to-noise ratio $\SNR = 10\log_\mathrm{10}\big((\frac{1}{M} \| \sum_{k \in \mathcal{K}} \tilde{\alpha}_k \V{s}(\tilde{\V{\psi}}_k)\|^2 + P B)/\sigma^2 \big)$.

\subsubsection{Empirical Substantiation of the ``Near-orthogonality'' Assumption}\label{ssec:OrthAss}

This study presents empirical evidence supporting Assumption 2. We consider a synthetic channel with a single \ac{sc}, i.e.,  
\eqref{eq:stack_recvgen} with $K=1$. The used settings are as follows: $N=27$, i.e. $\Delta=59.26$\,MHz, and the array has dimension $3 \times 3$ with $2$\,cm inter-element spacing, i.e.~$M=9$. Our algorithm uses a fixed threshold $\kappa^\ast=4$ and $L=50$.  
The dispersion parameters of the \ac{sc} are selected as follows: its delay is fixed to $\tilde{\tau}_1=\tilde{\tau}=10$\,ns
and its angle $\tilde{\varphi}_1=\tilde{\varphi}$ is drawn uniformly over $[0,2\pi)$ for each trial and independently across trials. The respective powers of noise, the \ac{dc} and the \ac{sc} are set such that $\SDR=-5$\,dB and $\SNR=20$\,dB and $\V{Q}$ is computed using \eqref{eq:Cov_approx}. 

To substantiate the ``near-orthogonality'' property claimed in Assumption 2  in each trial the cross-correlation coefficient $\V{s}(\hat{\V{\psi}}_j)^\text{H}\V{Q}^{-1} \V{s}(\hat{\V{\psi}}_i)/{\big(\V{s}(\hat{\V{\psi}}_i)^\text{H}\V{Q}^{-1} \V{s}(\hat{\V{\psi}}_i) \V{s}(\hat{\V{\psi}}_j)^\text{H}\V{Q}^{-1} \V{s}(\hat{\V{\psi}}_j)\big)}^{\frac{1}{2}}$ is calculated for any pair $(i,j)$ of indices of \acp{sc} detected by the algorithm and the mean and variance of these figures are obtained. The latter quantities are averaged over the $1000$ trials to yield $0.0123$ and $0.0401$, respectively. As a note the average number of detected \acp{sc} is $9.52$. We also plotted (not reported here due to space constraints) the estimated dispersion vectors of the detected \acp{sc} in their domain $\Psi$. 
By visual inspection we could qualitatively observe that vectors located outside an elliptically shaped region centered at the dispersion vector $[\tilde{\tau} \iist\iist \tilde{\varphi}]$ of the active \ac{sc} look uniformly distributed. The main axes of the boundary ellipse are set equal to 5 times the root-\acp{crb} for the estimation of the delay and angle. This choice ensures that estimated dispersion vectors located outside the elliptical region are very unlikely ($0,0001$) to be a noisy estimate of $[\tilde{\tau} \iist\iist \tilde{\varphi}]$.

\begin{figure*}[!t]
	\vspace*{-2mm}
	\captionsetup[subfigure]{captionskip=-0.1mm}
	\subfloat[\label{fig:WB_UWB_mean}]{\includegraphics[width=0.32\textwidth]{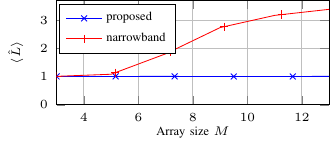}}\hspace*{2mm}
	\subfloat[\label{fig:WB_UWB_distance_rmse}]{\includegraphics[width=0.32\textwidth]{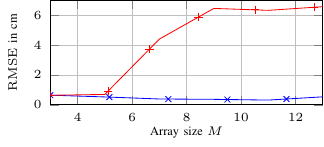}}\hspace*{2mm}
	\subfloat[\label{fig:WB_UWB_aoa_rmse}]{\includegraphics[width=0.32\textwidth]{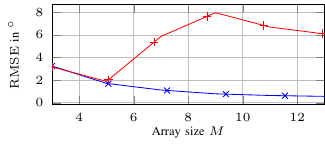}}
	\vspace*{-1mm}%
	\caption{Wideband versus narrowband detection and estimation of \acp{sc} in the \ac{awgn} channel: 
	\protect\subref{fig:WB_UWB_mean} mean number of detected \acp{sc}, \protect\subref{fig:WB_UWB_distance_rmse} \ac{rmse} of the distance estimates of detected \acp{sc}, and \protect\subref{fig:WB_UWB_aoa_rmse} \ac{rmse} of their angle estimates obtained with the proposed algorithm (in blue with crosses) and its narrowband version (in red with pluses) as a function of the array size $M$.}
	\label{fig:WB_UWB}
	\vspace{-4mm}
\end{figure*}

\subsubsection{Detection and Estimation of a Single \ac{sc}}\label{ssec:single_smc}

In this study, we first validate empirically the expression in \eqref{eq:PAtheorem1} as an approximation of the probability of detecting spurious \acp{sc} as well as an expression approximating the probability of not detecting an active \ac{sc} that we introduce now.
In a single-\ac{sc} scenario, the probability of missed detection can be approximated in the asymptotic regime $NM \rightarrow \infty$ by  $\mathbb{P}[\vert x({\hat{\V{\psi}}})\vert^2 < \kappa]$, where $\hat{\V{\psi}}$ is the estimated dispersion vector of the detected \ac{sc}. In this regime the distribution of $2\vert x(\hat{\V{\psi}})\vert^2$  
can be approximated by a non-central $\chi^2$ distribution with 2 degrees of freedom and non-centrality parameter $2\eta = 2\frac{|\tilde{\alpha}|^2}{\sigma^2} \sum_{m \in \mathcal{M}} 
\V{s}_m(\tilde{\V{\psi}})\M{Q}(\V{\eta}) \V{s}_m(\tilde{\V{\psi}})$ \cite{ShutinArxiv2015}. 
Making use of this result, the probability of missed detection in a single-\ac{sc} scenario 
is approximated by \cite{ShutinArxiv2015}  
\vspace*{-1mm}
\begin{align}
 \label{eq:PMtheorem1}
 P_\text{m}(\kappa) =  \int_0^{\kappa}\mathrm{e}^{-(x + \eta)}I_0(2\sqrt{\eta x}) \mathrm{d}x\ist.
 \\[-6mm]\nn
\end{align}

To numerically assess the accuracy of using \eqref{eq:PA} and \eqref{eq:PMtheorem1} as approximations of the probabilities of detecting spurious \acp{sc} and missing an active \ac{sc}, respectively, we modify the settings of the simulation scenario in Subsection~\ref{ssec:OrthAss} as follows: The array has size $5 \times 2$; $N=54$; 
$L=10$; $\SNR=\{5, 10, 20\}$\,dB; The threshold $\kappa^\ast$ of our algorithm
is a varying parameter. Other not explicitly mentioned settings stay as described in Subsection~\ref{ssec:OrthAss}. 
Note that by keeping the delay of the \ac{sc} fixed the non-centrality parameter stays constant and equal to $\eta=\{8.2, 12.2, 16.7\}$\,dB corresponding to the three $\SNR$ values.

Fig.~\ref{fig:PaPm} shows a comparison of $P_\text{f}(\kappa)$ in \eqref{eq:PA} (dashed lines) and $P_{\mathrm{m}}(\kappa)$ in \eqref{eq:PMtheorem1} (dash dotted lines) with the relative frequencies of, respectively, detecting a spurious \ac{sc} (solid lines) and missing the active \ac{sc} (dotted lines) computed from $1000$ trials. To compute the latter quantities we count the occurrence of two events that we now define. First we specify a rectangular region in $\Psi$ centered at the dispersion vector of the active \ac{sc} and with sides equal to $5$ times the square root of the respective \acp{crb} \cite{WildingACSSS2018}. The event ``false detection'' occurs if the estimated dispersion vector of at least one detected \ac{sc} lies outside the region. The event ``missed detection'' occurs if no \ac{sc} is detected or the estimated dispersion parameters of all detected \acp{sc} lie outside the region. 
The study is conducted under two assumptions on the covariance matrix $\V{Q}$ in \eqref{eq:C-EN} used in the generative model: $\V{Q}$ has the simplified form \eqref{eq:Cov_approx} (Fig.~\ref{fig:PaPm_knownC} and Fig.~\ref{fig:PaPm_unknownC}) and $\V{Q}$ has the general form \eqref{eq:C-EN} (Fig.~\ref{fig:PaPm_UWBC}).\footnote{We remind the reader that the algorithm is designed based on the simplified form \eqref{eq:Cov_approx}.}  
Furthermore, under the first assumption we distinguish between the two cases where the matrix $\V{Q}$ is known (Fig.~\ref{fig:PaPm_knownC}) and unknown (Fig.~\ref{fig:PaPm_unknownC}) to the algorithm, and thus is estimated in the latter case.

We see in Fig.~\ref{fig:PaPm_knownC} and Fig.~\ref{fig:PaPm_unknownC} that when $\V{Q}$ used in the generative model matches that used in the design of the algorithm, whether the algorithm knows or does not know said matrix has little impact on its performance. Fig.~\ref{fig:PaPm_UWBC} shows that when there is a mismatch, it only marginally affects the performance of the algorithm. Specifically, a comparison with Fig.~\ref{fig:PaPm_unknownC} shows that at large SNR values the number of spurious \acp{sc} is slightly increased due to the mismatch.

\subsubsection{Wideband versus Narrowband Detection and Estimation of \acp{sc} in the \ac{awgn} Channel}

State-of-the-art detection and estimation schemes are traditionally designed based on the narrow-band assumption, which  neglects the second occurrence of $g(\varphi, \V{p}_m)$ in \eqref{eq:s_signal} \cite{Fleury1999,RichterPhD2005}.
In this study we show that neglecting this term in the proposed algorithm leads to an increase of the number of detected spurious \acp{sc} as the size of the array increases. To quantitatively assess this effect, we modify the simulation scenario in Subsection~\ref{ssec:single_smc} as follows: the array is linear, i.e. $1\times M$, with inter-element spacing of $2$\,cm; $N=27$; $L=50$; $\SNR=20$\,dB; $P=0$ (\ac{wgn} only); $\epsilon=10^{-2}$;
$\sigma^2$ is assumed known. In this study and the subsequent ones we adopt the widely used convention in the radar community to convert (propagation) delays in their corresponding equivalent (propagation) distances. Fig.~\ref{fig:WB_UWB} depicts results obtained from $1000$ simulation trials that illustrate the behavior of the proposed algorithm (in blue with crosses) and of a simplified (narrowband) version of it that neglects the second occurrence of $g(\varphi, \V{p}_m)$ in \eqref{eq:s_signal} (in red with pluses) as a function of the array size $M$.
Fig.~\ref{fig:WB_UWB_mean}, Fig.~\ref{fig:WB_UWB_distance_rmse} and \ref{fig:WB_UWB_aoa_rmse} 
report respectively the mean number of detected \acp{sc}, the \ac{rmse} of the distance estimates, and the \ac{rmse} of the angle estimates. When $M$ is increased beyond $5$, the narrowband assumption is violated and the narrowband version of the algorithm detects additional spurious \acp{sc} with dispersion vectors located in the vicinity of that of the active \ac{sc}. 
By contrast, the \acp{rmse} achieved with the proposed algorithm decrease slightly as $M$ is increased.\footnote{To mitigate the impact of spurious \acp{sc} caused by large noise deviations, only detected \acp{sc} with distance and angle less than respectively $20$\,cm and $20\,^\circ$ apart of those of the active \ac{sc} are considered in the computation of the \ac{rmse} values.}

\subsubsection{High-resolution Capability of the Proposed Algorithm} \label{sec:overlapping_AWGN}

\begin{figure*}[!t]
	\centering
	\captionsetup[subfigure]{captionskip=-0.1mm}
	\subfloat[\label{fig:2D_K_10dB}]{\includegraphics[width=0.15\textwidth,height=0.13\textwidth]{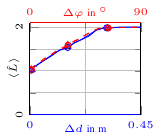}}\hspace*{-1.3mm}
	\subfloat[\label{fig:2D_P_10dB}]{\includegraphics[width=0.15\textwidth,height=0.13\textwidth]{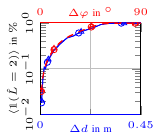}}\hspace*{-1.3mm}
	\subfloat[\label{fig:2D_tau_10dB}]{\includegraphics[width=0.23\textwidth,height=0.13\textwidth]{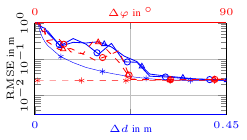}}\hspace*{-1.3mm}
	\subfloat[\label{fig:2D_phi_10dB}]{\includegraphics[width=0.23\textwidth,height=0.13\textwidth]{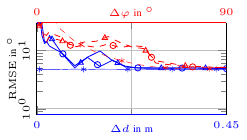}}\hspace*{-1mm}
	\subfloat[\label{fig:2D_alpha_10dB}]{\includegraphics[width=0.23\textwidth,height=0.13\textwidth]{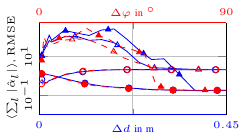}}\\
	\vspace*{-4mm}
	\subfloat[\label{fig:2D_K_30dB}]{\includegraphics[width=0.15\textwidth,height=0.13\textwidth]{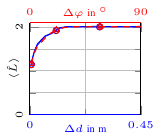}}\hspace*{-1.3mm}
	\subfloat[\label{fig:2D_P_30dB}]{\includegraphics[width=0.15\textwidth,height=0.13\textwidth]{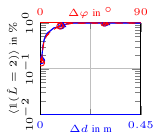}}\hspace*{-1.3mm}
	\subfloat[\label{fig:2D_tau_30dB}]{\includegraphics[width=0.23\textwidth,height=0.13\textwidth]{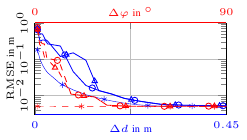}}\hspace*{-1.3mm}
	\subfloat[\label{fig:2D_phi_30dB}]{\includegraphics[width=0.23\textwidth,height=0.13\textwidth]{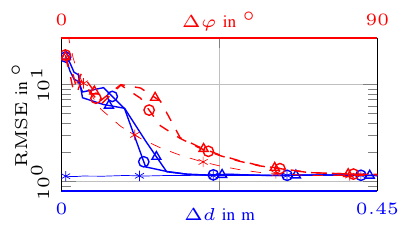}}\hspace*{-1.3mm}
	\subfloat[\label{fig:2D_alpha_30dB}]{\includegraphics[width=0.23\textwidth,height=0.13\textwidth]{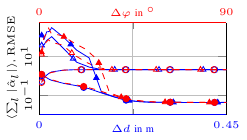}}\\
	\vspace*{-4mm}
	\subfloat[\label{fig:2D_K_50dB}]{\includegraphics[width=0.15\textwidth,height=0.13\textwidth]{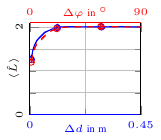}}\hspace*{-1.3mm}
	\subfloat[\label{fig:2D_P_50dB}]{\includegraphics[width=0.15\textwidth,height=0.13\textwidth]{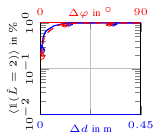}}\hspace*{-1.3mm}
	\subfloat[\label{fig:2D_tau_50dB}]{\includegraphics[width=0.23\textwidth,height=0.13\textwidth]{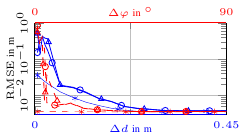}}\hspace*{-1.3mm}
	\subfloat[\label{fig:2D_phi_50dB}]{\includegraphics[width=0.23\textwidth,height=0.13\textwidth]{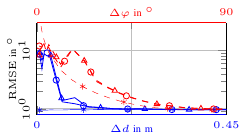}}\hspace*{-1.3mm}
	\subfloat[\label{fig:2D_alpha_50dB}]{\includegraphics[width=0.23\textwidth,height=0.13\textwidth]{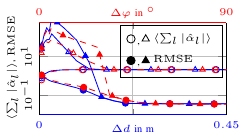}}\\
	\vspace*{2mm}
	\includegraphics{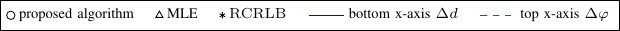}
	\vspace*{-1.5mm}
	\caption{Detection and estimation of two closely spaced \acp{sc} in the scenario depicted in Subsubsec.~\protect\ref{sec:overlapping_AWGN}.
	Blue solid and red dashed lines refer to the bottom and top horizontal-axis, respectively. The panels in each row depict results obtained for the same $\textrm{SNR}$ value, namely $\textrm{SNR}=10, 30, 50$\,dB, from the upper to the lower row. Panels in the columns depict from left to right the mean number of detected \acp{sc}, the relative frequency that exactly two \acp{sc} are detected, the \ac{rmse} of the estimated delays, the \ac{rmse} of the estimated angles, and the mean (unfilled) and \ac{rmse} (filled) of the moduli of the estimated amplitudes.
	}
	\label{fig:2D_unknownC}
	\vspace*{-1mm}
\end{figure*}

To study the super-resolution capability of the proposed algorithm, we consider a scenario with $K=2$ \acp{sc} with a controlled separation of their respective dispersion vector. The parameter vector $\tilde{\V{\psi}}_1$ of the first \ac{sc} is drawn uniformly over $\Psi$. 
The parameter vector of the second \ac{sc} is then set to either $\tilde{\V{\psi}}_2 = \tilde{\V{\psi}}_1 + [\Delta d/c \iist\iist 0]$ or $\tilde{\V{\psi}}_2 = \tilde{\V{\psi}}_1 + [0\iist\iist \Delta\varphi]$. 
The spacings $\Delta d$ and $\Delta\varphi$ are fractions of the \ac{rrl} in, respectively, distance ($c/\tilde{B}\approx 0.3$\,m) and angle ($56^\circ$) \cite{VanTrees2002}.
The complex amplitudes of both \acp{sc} have unit magnitude and their respective phases are drawn uniformly and independently. 
Other system parameters are set as follows: the array dimension is $3\times3$ with $2$\,cm inter-element spacing; 
$\SDR=6$\,dB; $\SNR=\{10, 30, 50\}$\,dB; $N=54$; $\epsilon=10^{-3}$; $L=10$.

In Fig.~\ref{fig:2D_unknownC}, we compare the performance of the proposed algorithm (circles) with a \ac{mle} algorithm (triangles) inspired by \cite{RichterPhD2005}. The modification consists in adapting the algorithm to our application scenario, adopting the same scheduling and the same thresholding as in our algorithm. Apart from the used scheduling, the modified algorithm strongly resembles that in \cite{NadlerTSP2011} with the information criterion adapted to our scenario. 
Performance versus spacings $\Delta d$ and $\Delta\varphi$ are depicted as, respectively, blue solid and red dashed curves.

The first two columns of Fig.~\ref{fig:2D_unknownC} present the mean number of detected \acp{sc} $\langle \hat{L} \rangle$ and the relative frequency that exactly two \acp{sc} are detected $\langle \mathds{1}(\hat{L}=2) \rangle$ versus spacing in distance and angle, respectively. Both algorithms are able to reliably find the correct number of \acp{sc} provided the respective dispersion vectors of the two \acp{sc} are sufficiently apart. At high $\SNR$ (see Fig.\,\ref{fig:2D_unknownC} \subref{fig:2D_K_50dB} to \subref{fig:2D_alpha_50dB}) the spacing values beyond which this occurs is as low as $0.15$\,m or $20\,^\circ$ for the system setting used in the study. 
Given that for this setting the \ac{rrl} in distance is $0.3$\,m and that in angle is $56^\circ$, this result demonstrates the superresolution capability of the proposed algorithm. At lower $\SNR$ (see Fig.\,\ref{fig:2D_unknownC} \subref{fig:2D_K_10dB} to \subref{fig:2D_alpha_10dB}) these values rise towards the \acp{rrl}. Note that these values are not only influenced by \ac{awgn} but also by the \ac{dc}, meaning that at high $\SNR$ the resolution capability of the algorithms is restricted by the $\SDR$. Further worth mentioning is that both algorithms tend to underestimate the number ($K=2$) of active \acp{sc} when the spacing is reduced. 
Columns three and four of Fig.~\ref{fig:2D_unknownC} depict the \ac{rmse} of, respectively, the distance and \ac{aoa} estimates, provided exactly two~\acp{sc} are detected. We associate the two detected \acp{sc} with the true \acp{sc} by means of the \ac{ospa} metric \cite{Schuhmacher2008}. To be able to use the  metric we normalize the estimated distances and angles with the \ac{rrl} in distance and in angle, respectively. The root of the sum of the \acp{crb} (delay and angle) of the two \acp{sc} are also depicted (lines with stars).
The estimates returned by both algorithms approach their respective root \ac{crb}, provided the spacing of the two \acp{sc} in $\Psi$ is large enough. 
The last column in Fig.~\ref{fig:2D_unknownC} presents the mean of the absolute value of the complex amplitudes $\langle \sum_l |\hat{\alpha}_l| \rangle$ (empty markers) and the \ac{rmse} of the absolute value of the complex amplitudes (filled markers) again provided exactly two \acp{sc} are detected. The two algorithms perform similarly in estimating the dispersion parameters; however our algorithm outperforms the modified \ac{mle} algorithm in estimating the complex amplitudes when the two \acp{sc} are closely spaced. 
This distinct behavior results from the specific structures of the algorithms: the modified \ac{mle} algorithm computes a least-squares estimate of the amplitudes, while the proposed algorithm computes a linear \ac{mmse} estimate based on the hyperparameter estimates, see \eqref{eq:HatMAlpha} and \eqref{eq:HatVAlpha}. 
When the estimated dispersion vectors of the two detected \acp{sc} are closely spaced, the least-squares estimator computes the inverse of an ill-conditioned matrix, while the linear \ac{mmse} estimator regularizes this matrix.

\vspace*{-2mm}
\subsection{Measured Radio Channels}\label{sec:MeasData}

\tikzsetnextfilename{Picture}
\begin{figure}
	\centering
	\includegraphics{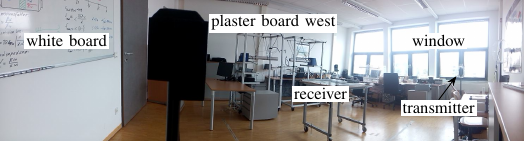}
	\vspace*{-5mm}
	\caption{Picture of the investigated room with the \ac{rx}, the \ac{tx} and some large-scale items labeled. A floorplan of the room is given in Fig.~7 in \cite[Appendix~B]{GreLeiFleWit:Arxiv2023}.}
	\label{fig:pic}
\end{figure}

\begin{figure*}[!t]
	\centering
	\vspace*{-2mm}
	\captionsetup[subfigure]{captionskip=-0.1mm}
	\subfloat[\label{fig:DAPS_original}]{\includegraphics[width=0.49\textwidth,height=0.14\textwidth]{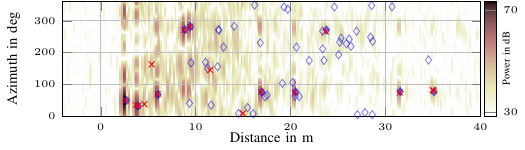}}\hspace*{-1mm}
	\subfloat[\label{fig:DPS}]{\includegraphics[width=0.49\textwidth,height=0.135\textwidth]{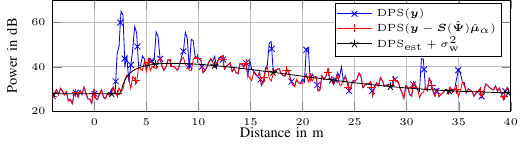}}\\[-0.1mm]
	\subfloat[\label{fig:DAPS_residual}]{\includegraphics[width=0.49\textwidth,height=0.14\textwidth]{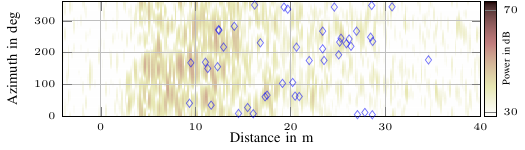}}\hspace*{-1mm}
	\subfloat[\label{fig:APS}]{\includegraphics[width=0.49\textwidth,height=0.135\textwidth]{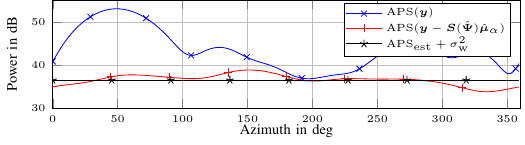}}
	\vspace*{-1mm}
	\caption{Selected application of the proposed algorithm to assess the dispersion characteristics (in delay and angle of arrival) of the channel from \ac{uwb} \ac{simo} measurement data: \protect\subref{fig:DAPS_original} estimated \ac{daps} and dispersion vectors of detected \acp{sc}, \protect\subref{fig:DPS} various estimated \ac{dps}, \protect\subref{fig:DAPS_residual} residual \ac{daps} and dispersion vectors of un-associated \acp{sc}, \protect\subref{fig:APS} various estimated \ac{aps}. Detailed descriptions of the measurement setting and the depicted results are given in Subsec.~\ref{sec:MeasData}.}
	\label{fig:DAPS}
	\vspace*{-5mm}
\end{figure*}

For the experimental study we used a channel sounding equipment that transmits an m-sequence of $7$\,GHz bandwidth at $6.95$\,GHz carrier frequency. Details about the equipment can be found in \cite{MeasureMINT2013}. After applying standard pre-processing steps (subtracting the cross-talk and equalizing the system response), the resulting signal is input to a filter with a root-raised-cosine transfer function with roll-off factor $0.6$ and bandwidth $B=1.6$\,GHz centered at $f_\mathrm{c}=6$\,GHz. The output signal with reduced bandwidth $B$ is then Fourier transformed and sampled over $[-B/2,+B/2]$ with frequency spacing $\Delta =6.8085\,$MHz to produce a length $N=235$ vector collecting these samples. 
A virtual $3\times 3$ antenna array with $2$\,cm inter-element spacing is emulated by means of a single antenna mounted on a positioning table. Since the received signals are essentially noise-free, \ac{wgn} was artificially added with power set such that $\SNR=40$\,dB to emulate the model in \eqref{eq:rx_signal}. The algorithm uses the following settings: $L=50$; $\epsilon=10^{-2}$.

The room where the measurements were performed is depicted in Fig.~\ref{fig:pic}. Based on a layout of it, see Fig.~7 in \cite[Appendix~B]{GreLeiFleWit:Arxiv2023}, a classical mirror source method \cite{Molisch2005WC} computes the positions of predicted virtual sources associated with rays from the Tx antenna to the center of gravity of the Rx array positions that undergo up to 5 reflections on walls or large objects (windows, boards). To each such predicted source corresponds a predicted \ac{sc} with dispersion vector computed from the position of the source, see details in \cite[Appendix~B]{GreLeiFleWit:Arxiv2023}.

The following analysis concerns measurements obtained with \ac{rx} position $\V{p}_1$ depicted in Fig.~7 in \cite[Appendix~B]{GreLeiFleWit:Arxiv2023}. Fig.~\ref{fig:DAPS}\subref{fig:DAPS_original} depicts the estimated \ac{daps} computed from the received signal \cite{Molisch2005WC}. Note that this power spectrum incorporates the smoothing function of the aperture of the measurement equipment \cite{Johnson1993}. This will be the case for all power spectra considered in this study. The red crosses and blue diamonds mark the estimated dispersion vectors of the \acp{sc} detected by the algorithm and the predicted \acp{sc}, respectively. To each detected \ac{sc} we associate (possibly no, one, or more than one) predicted \ac{sc} as follows. A predicted \ac{sc} is associated to a detected \ac{sc} if their respective distances and angles are no more apart than, respectively, $10$\,cm ($1/3$ of the \ac{rrl}) and $5\,^\circ$ ($1/10$ of the \ac{rrl}), see \cite[Appendix~B.A]{GreLeiFleWit:Arxiv2023} for the rationale behind this choice.
To most of the detected \acp{sc}, a unique predicted \ac{sc} is associated in this way. The two detected \acp{sc} with dispersion parameters $35$\,m and $80^\circ$ are associated with the same predicted \ac{sc}. An association could not be made for four detected \acp{sc}.
Fig.~\ref{fig:DAPS}\subref{fig:DAPS_residual} shows the estimated \ac{daps} computed from the residual signal $\V{y}-\V{S}(\hat{\V{\Psi}})\hat{\V{\mu}}$. It also includes the dispersion vectors of detected \acp{sc} (blue triangles) that could not be associated with any detected \ac{sc}. Clearly, the strong peaks in the estimated \ac{daps} depicted in Fig.~\ref{fig:DAPS}\subref{fig:DAPS_original} have vanished.
Fig.~\ref{fig:DAPS}\subref{fig:DPS} shows the estimated \ac{dps} computed from the original signal (solid blue with crosses) and from the residual signal (solid red with pluses), as well as the theoretical \ac{dps} of the \ac{dc} given in \eqref{eq:dps_parametric} with $\V{\eta}=\hat{\V{\eta}}$ plus the estimated noise variance $\hat{\sigma}$ (solid black with stars). The first two \ac{dps} result from averaging the respective \ac{daps} over the angle domain. The \ac{dps} obtained from the residual and reconstructed signals match well. This empirically justifies our choice of model \eqref{eq:dps_parametric}. 
Finally, Fig.~\ref{fig:DAPS}\subref{fig:APS} depicts the estimated \ac{aps} computed in a similar way as the \ac{dps} depicted in Fig.~\ref{fig:DAPS}\subref{fig:DPS}. The first two estimated \ac{aps} are obtained by averaging the respective \ac{daps} over the delay domain. It can be seen that the estimated \ac{aps} of the residual signal is nearly constant over the angle domain.
	
	\section{Conclusions}\label{sec:concl}
	In this paper, we derive and analyze a super-resolution algorithm for detecting and estimating specular components as well as estimating the power spectrum of the diffuse component plus noise in an ultra-wide band \acf{simo} multipath channel. Estimated parameters are among others the delay, angle-of-arrival, and complex amplitude of the detected specular components as well as the parameters of a parametric model of the delay power spectrum characterizing the diffuse component. The design of the algorithm is inspired by sparse Bayesian learning. As a result it embodies a pruning condition that determines whether a candidate specular component is considered active or not. The threshold of the pruning condition is adapted to control the probability of detecting spurious specular components.

Numerical studies in a synthetic environment show that the simplifying assumptions underlying the derivation of the algorithm are realistic and that the relative frequencies of detecting spurious specular components and missing active specular components are close to the respective probabilities derived theoretically. These studies also demonstrate several virtues of the algorithm: (a) its ability to still detect and accurately estimate specular components, even when their separation in delay and azimuth is down to half the Rayleigh resolution limit of the equipment; (b) it is robust in the sense that it tends to detect no more specular components than the actual ones. An experimental study illustrates the ability of the proposed algorithm to accurately infer the dispersive characteristics (in delay and angle of arrival) of the \ac{uwb} \ac{simo} channel. Owing to his high efficiency the proposed algorithm has promising potential applications in all aspects of wireless communications that exploit extended channel state information, such as \acf{isac} and radio-based localization.
	
	\appendices
	
	\section{Covariance of the 2-D $\chi^2$ Random Field $2|x(\V{\psi}_\cdot)|^2$}\label{app:covChi2}
	\begin{figure*}[!h]
	\begin{align}
		\mathbb{E}\Big[\frac{\partial x({\V{\psi}})\partial  x({\V{\psi}})^*}{\partial{\psi}_i \partial {\psi}_j }\Big] &= 
		2\Bigg(\frac{\sum\limits_{m \in \Set{M}}\Big(\frac{\partial \V{s}_m(\V{\psi})}{\partial{\psi}_j}\Big)^\text{H}\tilde{\M{Q}}^{-1}\frac{\partial \V{s}_m(\V{\psi})}{\partial{\psi}_i}}{\sum\limits_{m \in \Set{M}} \V{s}_m(\V{\psi})^\text{H}\tilde{\M{Q}}^{-1}\V{s}_m(\V{\psi})}
		- \frac{\sum\limits_{m \in \Set{M}} \Big(\frac{\partial \V{s}_m(\V{\psi})}{\partial{\psi}_i}\Big)^\text{H}\tilde{\M{Q}}^{-1} \V{s}_m(\V{\psi})  \Re\Big\{\sum\limits_{m' \in \Set{M}} \Big(\frac{\partial \V{s}_{m'}(\V{\psi})}{\partial{\psi}_j}\Big)^\text{H}\tilde{\M{Q}}^{-1}\V{s}_{m'}(\V{\psi})\Big\}}{\Big|\sum\limits_{m \in \Set{M}} \V{s}_m(\V{\psi})^\text{H}\tilde{\M{Q}}^{-1}\V{s}_m(\V{\psi})\Big|^2} \nn \\
		&\hspace{6mm}- \frac{\sum\limits_{m \in \Set{M}} \V{s}_m(\V{\psi})^\text{H}\tilde{\M{Q}}^{-1}\frac{\partial \V{s}_m(\V{\psi})}{\partial{\psi}_j} \Re\Big\{\sum\limits_{m' \in \Set{M}}  \Big(\frac{\partial \V{s}_{m'}(\V{\psi})}{\partial{\psi}_i}\Big)^\text{H}\tilde{\M{Q}}^{-1} \V{s}_{m'}(\V{\psi}) \Big\}}{\Big|\sum\limits_{m \in \Set{M}}  \V{s}_m(\V{\psi})^\text{H}\tilde{\M{Q}}^{-1}\V{s}_m(\V{\psi})\Big|^2} \nn \\
		&\hspace{6mm}+ \frac{\Re\Big\{\sum\limits_{m \in \Set{M}}\Big(\frac{\partial \V{s}_{m}(\V{\psi})}{\partial{\psi}_i}\Big)^\text{H}\tilde{\M{Q}}^{-1} \V{s}_{m}(\V{\psi}) \Big\} \Re\Big\{\sum\limits_{m' \in \Set{M}}\Big(\frac{\partial \V{s}_{m'}(\V{\psi})}{\partial{\psi}_i}\Big)^\text{H}\tilde{\M{Q}}^{-1} \V{s}_{m'}(\V{\psi}) \Big\}}{\Big|\sum\limits_{m \in \Set{M}}  \V{s}_m(\V{\psi})^\text{H}\tilde{\M{Q}}^{-1}\V{s}_m(\V{\psi})\Big|^2}
		\Bigg)
		\label{eq:derivativegeneral}
	\end{align}
	\hrulefill
\end{figure*}

The $(i,j)$-entry of \eqref{eq:cov_randomfield} is given in \eqref{eq:derivativegeneral} with $\tilde{\V{Q}}$ according to \eqref{eq:BDEC}. The partial derivatives of the signal $S(f; \tau, \varphi,\V{p}^{(m)}) = \text{e}^{j2\pi f_c g(\varphi,\V{p}_m)} \underline{S}(f) \text{e}^{-j2\pi f\tau}$ w.r.t. $\tau$ and $\varphi$ are, respectively, $\partial {\V{s}}_m(\V{\psi})/\partial \tau = -\mathrm{e}^{j 2 \pi f_\mathrm{c} g(\varphi, \V{p}_m)} \dot{\underline{\V{s}}}(\tau) $ and 
$\partial \V{s}_m(\V{\psi})/\partial \varphi = d_m({\varphi}) \exp(j2\pi f_\mathrm{c} g(\varphi, \V{p}_m)) j2\pi f_\mathrm{c} \underline{\V{s}}(\tau)$
with $d_m({\varphi}) = \partial g(\varphi, \V{p}_m))/\partial\varphi$ and $\dot{\underline{\V{s}}}(\tau) = \partial \underline{\V{s}}(\tau)/\partial \tau $. Due to the centro-symmetry of the spatial aperture, see Assumption~\ref{as:KnoCovM}, for any $m \in \mathcal{M}$, there exists an index $m' \in\mathcal{M}$ such that $\V{p}_{m'}-\V{p}=-(\V{p}_m-\V{p})$. As a result
\vspace*{-1mm}
\begin{align}
\sum_{m \in \Set{M}}d_m({\varphi}) = 0 \ist.
\label{eq:centerGrav}\\[-6mm]\nn
\end{align}

We now proceed with the computation of the specific entries of \eqref{eq:cov_randomfield}.

\subsubsection{Second-order Partial Derivatives of $x({\V{\psi}})$ w.r.t. $\tau$} One can easily check that because of \eqref{eq:centerGrav} the second and third terms of \eqref{eq:derivativegeneral} vanish in this case. 
As a result, we can write
\begin{align}
&\mathbb{E}\Big[\frac{\partial x({\V{\psi}})\partial x({\V{\psi}})^*}{\partial{\tau}^2 }\Big] \nn \\[-1mm]
&\hspace*{2mm}= 2\Bigg(\frac{\sum\limits_{m \in \Set{M}} \dot{\underline{\V{s}}}(\tau)^\text{H}\tilde{\M{Q}}^{-1} \dot{\underline{\V{s}}}(\tau)}{\sum\limits_{m \in \Set{M}} \underline{\V{s}}(\tau)^\text{H}\tilde{\M{Q}}^{-1}\underline{\V{s}}(\tau) } \rmv-\rmv \frac{\Re\Big\{\sum\limits_{m \in \Set{M}} \dot{\underline{\V{s}}}(\tau)^\text{H}\tilde{\M{Q}}^{-1} \underline{\V{s}}(\tau) \Big\}^2 }{\Big|\sum\limits_{m \in \Set{M}} \underline{\V{s}}(\tau)^\text{H} \tilde{\M{Q}}^{-1} \underline{\V{s}}(\tau)\Big|^2} \Bigg)  \nn\\[-1mm]
&\hspace*{2mm}= 2\frac{\big(\dot{\underline{\V{s}}}(\tau)^\text{H}\tilde{\M{Q}}^{-1}\dot{\underline{\V{s}}}(\tau)\big)^2}{\big(\underline{\V{s}}(\tau)^\text{H}\tilde{\M{Q}}^{-1}\underline{\V{s}}(\tau)\big)^2}\nn \\
&\hspace*{10mm}\times\Bigg(1  - \frac{\Re\big\{ \dot{\underline{\V{s}}}(\tau)^\text{H}\tilde{\M{Q}}^{-1} \underline{\V{s}}(\tau)\big\}^2 }{\big(\underline{\V{s}}(\tau)^\text{H}\tilde{\M{Q}}^{-1} \underline{\V{s}}(\tau)\big)^2 \big(\dot{\underline{\V{s}}}(\tau)^\text{H}\tilde{\M{Q}}^{-1} \dot{\underline{\V{s}}}(\tau)\big)^2} \Bigg)\nn\\
&= 8\pi^2 b(\tau) e(\tau) \label{eq:derivativetaufinal}
\end{align}
where
\begin{align}
b(\tau) = \big[\big(\dot{\underline{\V{s}}}(\tau)^\text{H}\tilde{\M{Q}}^{-1}\dot{\underline{\V{s}}}(\tau)\big)/\big(4\pi^2 \underline{\V{s}}(\tau)^\text{H}\tilde{\M{Q}}^{-1} \underline{\V{s}}(\tau)\big)\big]^{1/2}
\end{align}
and 
\begin{align}
	e(\tau) = 1  - \frac{\Re\big\{ \dot{\underline{\V{s}}}(\tau)^\text{H}\tilde{\M{Q}}^{-1} \underline{\V{s}}(\tau)\big\}^2  }{\big(\underline{\V{s}}(\tau)^\text{H}\tilde{\M{Q}}^{-1} \underline{\V{s}}(\tau)\big)^2 \big(\dot{\underline{\V{s}}}(\tau)^\text{H}\tilde{\M{Q}}^{-1} \dot{\underline{\V{s}}}(\tau)\big)^2}
\end{align}
is a delay-dependent loss factor that depends of the structure of the noise vector $\V{n}$. Note that if $\V{n}$ is white, $e(\tau) = 1$, otherwise $e(\tau) < 1$, typically.

\subsubsection{Second-order Partial Derivatives of $x({\V{\psi}})$ w.r.t. $\varphi$}
In this case, the last three terms in \eqref{eq:derivativegeneral} vanish, again because of  \eqref{eq:centerGrav}. We readily obtain,
\begin{align}
\mathbb{E}\Big[\frac{\partial x({\V{\psi}})\partial x({\V{\psi}})^*}{\partial{\varphi}^2 }\Big] = \frac{8\pi^2 f_{\mathrm{c}}^2}{M} \sum\limits_{m \in \Set{M}}d_m^2(\varphi)\ist.
\label{eq:derivativephi}
\end{align}
\subsubsection{Second-order Partial Derivatives of $x({\V{\psi}})$ w.r.t. $\tau$ and $\varphi$}
In this case \eqref{eq:centerGrav} make all terms in \eqref{eq:derivativegeneral} vanish. Thus,
\begin{align}
\mathbb{E}\Big[\frac{\partial x({\V{\psi}})\partial x({\V{\psi}})^*}{\partial{\tau}\partial{\varphi} }\Big] &= 0\ist.
\label{eq:derivativephitau}
\end{align}

	\section{Validation of the SMCs detected by the algorithm}\label{app:addres}

In this appendix, we provide a qualitative study that attempt to relate the \acp{sc} detected by the proposed algorithm to probable propagation mechanisms in the environment where the experimental data were collected. The results of this study supplement those presented in Subsec.~\ref{sec:MeasData}.

\begin{figure}[!t]
	\centering
	\vspace*{-2mm}
	\footnotesize
	\includegraphics[width=1\columnwidth]{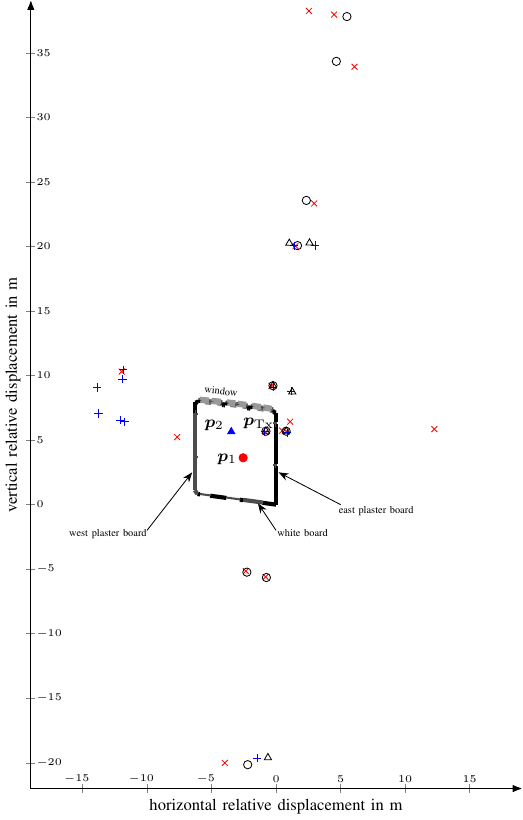} 		
	\vspace{-3mm}
	\caption{Floorplan of the investigated environment including the (fixed) \ac{tx} position $\V{p}_{\text{Tx}}$, the two selected positions $\V{p}_1$ (red filled circle) and $\V{p}_2$ (blue filled triangle) of (the center of gravity of) the \ac{rx} array, the locations of estimated (virtual) sources for \ac{rx} array position $\V{p}_1$ (red crosses) and \ac{rx} array position $\V{p}_2$ (blue pluses) and the positions of associated predicted sources (black circles and triangles, respectively).}
\label{fig:sbl_fp_real}
\end{figure}

A 2-D coordinate system including the layout of the room where the measurements were taken is shown in Fig.~\ref{fig:sbl_fp_real}. Also reported are the two selected positions $\V{p}_1$ and $\V{p}_2$ of (the center of gravity of) the \ac{rx} (virtual) array and the fixed position $\V{p}_{\text{Tx}}$ of the (single) \ac{tx} antenna. 
We recall that the mirror source method \cite{Molisch2005WC} computes the positions of predicted virtual sources associated with rays from the Tx antenna to the Rx array positions that undergo up to 5 reflections on walls or large objects (windows, boards). For the sake of conciseness we refer to virtual sources in the sequel as sources. The position, denoted by $\hat{\V{p}}_l$, of the predicted source corresponding to the $l$th \ac{sc}, $l=1,...\hat{L}$ detected by the algorithm is computed based on the estimated dispersion vector of the \ac{sc} using the relation $\hat{\V{p}}_l = \V{p} + c\hat{\tau}_l [\cos(\hat{\varphi}_l) \ist\ist \sin(\hat{\varphi}_l)]\trans$, 
where $\V{p}$ either equals $\V{p}_1$ or $\V{p}_2$. These  positions are depicted in Fig.~\ref{fig:sbl_fp_real} as red crosses and blue pluses for the \ac{rx} array positions $\V{p}_1$ and $\V{p}_2$, respectively.  

The procedure described next attempts to associates detected sources and predicted sources. Possibly no, one, or more than one predicted \acp{sc} are associated to each detected \ac{sc} as follows. A predicted \ac{sc} is associated to a detected \ac{sc} if their respective distances and angles are no more than, respectively, $10$\,cm ($1/3$ of the \ac{rrl} in distance) and $5\,^\circ$ ($1/10$ of the \ac{rrl} in angle) apart. These selected values are within the same order of magnitude as, respectively,  the  $5$\,cm approximate accuracy of the floorplan (measured with a tape measure) and the \acp{crb} of the estimated distances and angles.
The positions of successfully associated predicted sources are depicted in Fig.\,\ref{fig:sbl_fp_real} as black circles and triangles for \ac{rx} positions $\V{p}_1$ and $\V{p}_2$, respectively. 

The algorithm is able to identify the LOS, most of the predicted first-order reflections and some predicted higher-order reflections for both \ac{rx} positions. Worth noting are the rays with reflections up to order five via the white board and the window highlighted in Fig.\,\ref{fig:sbl_fp_real} and Fig.~\ref{fig:pic} and  the second-order rays with reflections via the west plaster board and the east plaster board. The two former items are made of more reflective materials than the two latter. 
Furthermore, scattering from a metallic frame (see Fig.~\ref{fig:pic}) that was not considered in the  mirror source method could explain the detected source located at approximately $[-8\iist\iist 6]$\,m close to the west plaster board. For \ac{rx} position $\V{p}_2$ many detected sources are found in a region around $[-15\iist\iist 8]$\,m, that are likely to originate from scattering from this metallic frame.

	\renewcommand{\baselinestretch}{0.9}\small\normalsize
	
	\bibliographystyle{IEEEtran}
	\bibliography{IEEEabrv,references}

\end{document}